\documentclass[prb,onecolumn,lengthcheck,superscriptaddress]{revtex4-1}

\usepackage{natbib}
\usepackage{graphicx,color}
\usepackage{amsmath,amssymb}
\usepackage{textcomp}
\usepackage{braket,slashed}
\usepackage{hyperref}
\usepackage[english]{babel}
\usepackage{braket}
\usepackage{soul}

\usepackage[mathscr]{euscript}
\usepackage{dsfont}

\newcommand{\operator}[1]{\ensuremath{\hat{#1}}}

\DeclareMathOperator{\order}{\mathscr{O}}

\newcommand{\Tr}{\ensuremath{\mathrm{Tr}}}
\renewcommand{\d}{\ensuremath{\mathrm{d}}}
\newcommand{\ic}{\ensuremath{\mathrm{i}}}
\newcommand{\ec}{\ensuremath{\mathrm{e}}}
\newcommand{\LL}{\ensuremath{\mathrm{L}}}
\newcommand{\rket}[1]{\ensuremath{|#1)}}
\newcommand{\rbra}[1]{\ensuremath{(#1|}}

\begin{document}
\title{Continuous matrix product states with periodic boundary conditions and an application to atomtronics}

\author{Damian Draxler}
\affiliation{Faculty of Physics, University of Vienna, Boltzmanngasse 5, A-1090 Wien, Austria}
\author{Jutho Haegeman}
\affiliation{Department of Physics and Astronomy, University of Ghent, Krijgslaan 281 S9, B-9000 Ghent, Belgium}
\author{Frank Verstraete}
\affiliation{Department of Physics and Astronomy, University of Ghent, Krijgslaan 281 S9, B-9000 Ghent, Belgium}
\affiliation{Faculty of Physics, University of Vienna, Boltzmanngasse 5, A-1090 Wien, Austria}
\author{Matteo Rizzi}
\affiliation{Institut f\"{u}r Physik, Johannes Gutenberg-Universit\"{a}t Mainz, Staudingerweg 7, D-55099 Mainz, Germany}

\begin{abstract}
We introduce a time evolution algorithm for one-dimensional quantum field theories with periodic boundary conditions. This is done by applying the Dirac-Frenkel time-dependent variational principle to the set of translational invariant continuous matrix product states with periodic boundary conditions. Moreover, the ansatz is accompanied with additional boundary degrees of freedom  to study quantum impurity problems. The algorithm allows for a cutoff in the spectrum of the transfer matrix and thus has an efficient computational scaling. In particular we study the prototypical example of an atomtronic system - an interacting Bose gas rotating in a ring shaped trap in the presence of a localised barrier potential.     
\end{abstract}

\maketitle

The last decade has shown a tremendous development of tensor network based methods for studying strongly correlated quantum many-body systems \cite{Orusreview}. 
More recently the  scope of these methods was extended by introducing the variational class of continuous matrix product states (cMPS) \cite{PhysRevLett.104.190405,PhysRevB.88.085118}, which allows to simulate one dimensional (1D) quantum field theories directly in the continuum.
Systems with continuous degrees of freedom such as superfluid atom circuits (i.e., cold atoms trapped in ring-shaped traps), have recently attracted a lot of attention as building blocks in the emerging field of \textit{atomtronics} \cite{PhysRevLett.106.130401,PhysRevLett.110.025302,HysteresisCamp,PhysRevA.75.023615,PhysRevA.82.063623,Beeler}.
While (contact-)interacting quantum particles in the 1D continuum with periodic boundaries provide prototypical examples of integrable systems \cite{PhysRev.130.1605,yang1969thermodynamics,Korepin}, the presence of controlled impurities to steer the atomtronic system often breaks the integrability conditions.
Analytical solutions can thus only be found in special limits \cite{PhysRevB.55.6551,PhysRevLett.89.096802}, and even numerical studies do suffer from an intrinsic approximation as they always rely on discretised versions of the a priori continuous system \cite{PhysRevLett.113.025301}. 
       
In this paper we present a time evolution algorithm for  translational invariant cMPS with periodic boundary conditions (pbc) which is perfectly well suited for the numerical study of atomtronic systems. The algorithm is based on the time-dependent variational principle (TDVP) \cite{TDVP1,TDVP2,TDVP3}, which was first introduced to the field of tensor network methods in Ref.~\onlinecite{PhysRevLett.107.070601} by applying it to the variational class of uniform matrix product states (MPS) in the thermodynamic limit and later to finite systems with open boundary conditions (obc) \cite{TDVPnew,PhysRevLett.111.207202}.
Only recently TDVP was also applied to cMPS with obc \cite{QGP,PhysRevLett.111.020402} (both for finite and infinite systems) but the extension of the algorithm to the pbc case has not been done so far neither for MPS nor for cMPS. The cMPS ansatz is further accompanied with additional boundary degrees of freedom to study Hamiltonians which contain local translational symmetry braking terms akin those in quantum impurity problems.

The difficulty in studying systems with pbc with tensor networks in general stems from the fact that contracting a tensor network with pbc is numerically much more costly than contracting one with obc. In the spirit of Refs.~\onlinecite{PhysRevLett.93.227205,PhysRevB.81.081103,PhysRevB.83.125104} it is however possible to overcome this difficulty by  allowing effectively for a cutoff in the spectrum of the corresponding transfer matrix of the network. We present an analog of this procedure for uniform cMPS  such that our algorithm has the same computational cost as conventional pbc DMRG methods  \cite{PhysRevB.81.081103}. 
Moreover we anticipate here that our method will not require the introduction of spatially dependent tensors, except for \emph{one} variational boundary matrix at the impurity position, therefore providing an approach (almost) as cheap as translationally invariant cMPS algorithms.

The paper is organized as follows. In section\ref{sec:secTDVP} the algorithm is introduced and we explain the aforementioned cutoff procedure whereas in section\ref{sec:secModel} we define the Hamiltonian. In section\ref{sec:secResults} we present the results which are extensively benchmarked with DMRG based lattice simulations and classical Gross-Pitaevskii solutions.    

 \subsection{Time-dependent variational principle}\label{sec:secTDVP}
 
We start by defining a uniform cMPS with pbc and bond dimension $D$ as
\begin{equation}\label{eq:cMPSfunc}
\ket{\Psi[Q,R,B]} = \Tr \Big[ B \mathscr{P}\ec^{\int_{0}^{\LL} Q \otimes \operator{1} + R \otimes \operator{\psi}^\dagger(x)\,\mathrm{d}x} \Big] \ket{\Omega}
\end{equation}
with $R, Q$ being matrices in $\mathbb{C}^{D\times D}$, $\mathscr{P}\ec$ the path-ordering exponential, $L$ the system size and $\ket{\Omega}$ the vacuum which is annihilated by the field operator $\operator{\psi}(x)$. The field operators are assumed to be bosonic satisfying the commutation relation $[\operator{\psi}(x),\operator{\psi}^{\dagger}(y)] = \delta (x-y)$ and the boundary condition $\operator{\psi}(0) = \operator{\psi}(L)$. We also include the boundary matrix $B \in \mathbb{C}^{D\times D}$ to the variational space which accounts for the possibility of having a localised barrier at position $x=0$ and for a possible phase twist in the boundary conditions.  The importance of this boundary matrix for cMPS simulations was already  discussed previously in the context of the relativistic Casimir effect \cite{PhysRevLett.105.251601} and in the derivation of the TDVP equations for systems with obc \cite{QGP}.  Moreover, its relation to the cMPS gauge theory in general was also discussed in Ref.~\onlinecite{PhysRevB.88.085118}. The invariance of a cMPS under multiplicative gauge transformations \cite{PhysRevB.88.085118} of the form $R \to gRg^{-1} , Q \to gQg^{-1} , B \to gBg^{-1}$ with $g\in {\rm{GL}}(\mathbb{C},D)$ allows to work with $Q$ and $R$ matrices s.t. $Q+Q^{\dagger} + R^{\dagger}R = 0$ which we refer to as the \textit{left gauge condition}. This condition is always satisfied if we set $Q = -1/2R^{\dagger}R - \ic K$ with $K$ being a hermitian matrix.    
\\
The TDVP was originally formulated as an action principle \cite{TDVP1,TDVP2} over a given (complex) variational manifold giving rise to the Euler-Lagrange equations. However, TDVP can also be interpreted as a projected time-dependent Schr\"{o}dinger equation $\ic\frac{\d}{\d t} \ket{\Psi} = \operator{P}_{\Psi}\operator{H}\ket{\Psi}$,  where $\operator{H}$ is a Hamiltonian and $\operator{P}_{\Psi}$ projects the  r.h.s. to the tangent space of the variational manifold \cite{GeometryMPS} at the point $\ket{\Psi}$.  This projector guarantees that the state will never leave the variational manifold which can be seen as the guiding principle behind TDVP, i.e. finding the optimal approximation to the true time evolution within a certain variational manifold. In our setting the variational manifold is the set of cMPS with fixed bond dimension $D$ and the tangent states spanning the corresponding tangent space are defined as \cite{PhysRevB.88.085118}
\begin{widetext}
\begin{eqnarray} \label{eq:tangent}
\ket{\Phi[V,W,Y]}  & := & 
\left[ Y \frac{\delta}{\delta B} + \int_{0}^{\LL}  \d x \; \left( V \frac{\delta}{\delta Q} + W \frac{\delta}{\delta R} \right) \right] \ket{\Psi[Q,R,B]}\nonumber \\
& = & \Tr \Big[ Y \operator{M}(0,\LL) \Big] + \int_{0}^{\LL}  \d x \;  \Tr \Big[ B \operator{M}(0,x) \Big(V \otimes \operator{1} 
+ W \otimes \operator{\Psi}^{\dagger}(x) \Big) \operator{M}(x,\LL) \Big]
\end{eqnarray}
\end{widetext}   
where $W$, $V$ ane $Y$ are variational matrices in $\mathbb{C}^{D\times D}$. We also introduced the notation $\operator{M}(x,y) = \mathscr{P}\ec^{\int_{x}^{y} \d z Q \otimes \operator{1} + R \otimes \operator{\psi}^\dagger(z)} $.

Deriving the TDVP equations now amounts to finding the optimal tangent vector that the exact Schr\"{o}dinger evolution is projected onto. By introducing a time-dependent parameterisation  $(Q(t),R(t),B(t))$ we therefore have $\ket{\Phi[\dot{Q},\dot{R},\dot{B}]} = -\ic\operator{P}_{\Psi}\operator{H}\ket{\Psi[Q,R,B]}$ (with the dot notation referring to time derivatives)  and  $\ket{\Phi[\dot{Q},\dot{R},\dot{B}]}$ is obtained by performing the minimisation 
\begin{equation*}
 \min_{(V,W,Y)} \|\ket{\Phi[V,W,Y]} + \ic\operator{H}\ket{\Psi[Q,R,B]}\|_{{\color{red}{2}}} \;.
 \end{equation*}
The solution to this minimisation problem is the optimal tangent vector given by $(\dot{Q},\dot{R},\dot{B})$ and obeys the following equation ( see e.g. \cite{TDVP3,PhysRevD.88.085030})
\begin{equation}\label{eq:TDVPeq}
\ic  \left[
\begin{array} {c}
\dot{Q} \\
\dot{R}  \\
\dot{B}
\end{array}
\right] = G^{-1} \left[ \begin{array} {c} g_{1} \\ g_{2} \\ g_{3}  \end{array} \right] \;. 
\end{equation}
This differential equation has then to be solved for $(Q,R,B)$ in order to find the optimal trajectory within our variational manifold. The Gram matrix $G$ (see Ref.~\onlinecite{PhysRevB.88.075133} and Ref.~\onlinecite{GeometryMPS} for a detailed discussion of the geometric properties of (c)MPS manifolds)  and the vector $[g_{1} ; g_{2} ; g_{3} ]$ are defined via
\begin{align}\label{eq:defTDVPeq}
&\braket{\Phi[\overline{V},\overline{W},\overline{Y}]|\Phi[V,W,Y]} = \left[V^{\dagger} W^{\dagger} Y^{\dagger}\right] G \left[ \begin{array} {c} V \\ W \\ Y \end{array} \right]  \\
\label{eq:defTDVPeq2}
&\bra{\Phi[\overline{V},\overline{W},\overline{Y}]}\operator{H}\ket{\Psi(Q,R,B)} = \left[V^{\dagger} W^{\dagger} Y^{\dagger}\right] \left[ \begin{array} {c} g_{1} \\ g_{2} \\ g_{3} \end{array} \right] \;. 
\end{align}
The  square brackets are  assumed to be vectors of length $3D^{2}$ and the Gram matrix $G$ is a hermitian matrix of size $(3D^{2}\times 3D^{2})$  and both of them are functions of $Q$, $R$ and $B$.  Note also that in the above expressions we made use of the trace property $\Tr[X^{\dagger}] = \Tr[\overline{X}]$ with some $X\in\mathbb{C}^{D\times D}$ and the bar notation means complex conjugation. Further we will refer to the r.h.s. of Eq.~\eqref{eq:TDVPeq} as the gradient. It is important to note that $G$  has a non-vanishing kernel of dimension $D^{2}$. This means that the inverse in Eq.~\eqref{eq:TDVPeq} has to be understood as a pseudo-inverse acting solely on the orthogonal complement of the kernel of $G$.   The existence of this kernel can be traced back to the fact that every tangent state $\ket{\Phi[V,W,Y]}$ is invariant under additive gauge transformations \cite{PhysRevB.88.085118}  of the form $W \to W+[X,R] ,V \to V+[X,Q]$ and $Y \to Y+[X,B]$ with some $X\in\mathbb{C}^{D\times D}$. From Eq.~\eqref{eq:defTDVPeq}, and the fact that the tangent space is a linear vector space, as visible from Eq.~(\ref{eq:tangent}), we can conclude that the states $\ket{\Phi[[X,Q],[X,R],[X,B]]}$ must have norm  zero and the corresponding vectors $[[X,Q];[X,R];[X,B]]$ span the null space of $G$. It is convenient to work in a gauge s.t. $V = -R^{\dagger}W$ which eliminates all $D^{2}$ gauge degrees of freedom and thereby maps $G$ to a $2D^{2}$-dimensional effective Gram matrix with full rank. Note that it is always possible to find an appropriate $X$ which brings any $(V,W)$ into this gauge provided that neither $R$ nor $Q$ are zero or the identity matrix.  Further, this choice of gauge allows for a \textit{left gauge} conserving update procedure as we will not update $Q$ but rather $K$ as $\ic\dot{K} = 1/2(R^{\dagger}\dot{R}-\dot{R}^{\dagger}R)$. This equation guarantees that $K$ will stay hermitian to first order in $dt$ and the corresponding $Q = -1/2R^{\dagger}R - \ic K$ will thus satisfy the left gauge condition throughout the time evolution.   

It is worth mentioning that in the context of TDVP the difference in the computational scaling between systems with obc and pbc is that for obc it is always possible, just like for standard MPS,  to find a gauge such that $G=\mathbb{I}$, which drastically reduces the cost of solving \eqref{eq:TDVPeq}, whereas for pbc  in general such a gauge does not exist.

When  solving Eq.~\eqref{eq:TDVPeq} in imaginary time (to find an approximation to the ground state of some Hamiltonian) it is important to restrict to tangent states that are orthogonal to $\ket{\Psi(Q,R,B)}$, i.e. $ \Braket{\Phi[\overline{V},\overline{W},\overline{Y}] | \Psi(Q,R,B)} = \left[V^{\dagger} W^{\dagger} Y^{\dagger}\right]  \left[ y_{1};y_{2};y_{3}\right] = 0$ \;.
This constraint is due to the fact that  $\ket{\Psi(Q,R,B)}$ itself is an element of the tangent space as $\ket{\Phi[\mathbb{I},0_{D\times D},0_{D\times D}]} = \LL\ket{\Psi(Q,R,B)}$. Hence, without imposing  orthogonality  the evolution would not preserve the norm of the state. Thus we not only have to project Eq.~\eqref{eq:TDVPeq} onto the orthogonal complement of the kernel of $G$ but also to the subspace orthogonal to the vector $[y_{1};y_{2};y_{3}]$. Note that for real time evolution orthogonality does not have to be imposed,  since then the TDVP equations are symplectic differential equations \cite{TDVP3} and as such  respect conservation of norm and energy. 
 
In this paper we will only focus on ground state properties hence solving Eq.~\eqref{eq:TDVPeq} in imaginary time. In the case of a simple Euler integration the imaginary time TDVP equations are equivalent to a steepest descend optimization where the cost function is the expectation value of the Hamiltonian, i.e., the energy of the state. Further improvement of the convergence properties can be achieved by  applying a version of the non-linear conjugate gradient search on a curved manifold. This is a rather involved task as the energy has to be minimised along geodesics and the parallel transport of tangent states corresponding to different tangent spaces has to be calculated in general. However, for small enough time steps the parallel transport can be neglected \cite{PhysRevD.88.085030}  and the only ingredients needed for a non-linear CG search are then the gradient and the metric  $G$ defined in Eq.~\eqref{eq:TDVPeq}. 
  
\emph{Cut off in the transfer matrix ---}
We will now proceed  by calculating the vector $[y_{1};y_{2};y_{3}]$ for a general state given by $(Q,R,B)$ and thereby explain the neccessary \textit{cut off} procedure to make the algorithm numerically efficient.  This vector will be a function of $(Q,R,B)$ and only enters the algorithm as an orthogonality constraint which can even be ignored in the case of real-time evolution.  By using standard cMPS calculus techniqus \cite{PhysRevB.88.085118} we find that
\begin{align}\label{eq:gsortho}
&\Braket{\Phi[\overline{V},\overline{W},\overline{Y}] | \Psi(Q,R,B)}  = \nonumber \\ 
&\int_{0}^{\LL} \d x \Tr \Big[ B\otimes\overline{B} \ec^{xT}\Big(\mathbb{I}\otimes\overline{V} + R\otimes\overline{W} \Big)\ec^{(\LL-x)T} \Big]   \nonumber \\ 
&+ \Tr \Big[ B\otimes\overline{Y} \ec^{\LL T}\Big]
\end{align}
 where $T=Q\otimes\mathbb{I} + \mathbb{I}\otimes\overline{Q}+R\otimes\overline{R}$ is the generator of the cMPS transfer matrix $E(x,y) = \ec^{(y-x)T}$. The matrix $T$ can always be rescaled \cite{PhysRevB.88.085118} to have one zero eigenvalue, with all other eigenvalues having non-positive real parts:
we notice that this property is automatically fulfilled by virtue of the left gauge condition, $Q+Q^\dagger + R^\dagger R = 0$.
We also assume that these eigenvalues are ordered with decreasing real part such that $|\ec^{\lambda_{1}}| \geq |\ec^{\lambda_{2}}| \geq \cdots \geq |\ec^{\lambda_{D^{2}}}|$. If we insert a spectral decomposition of $T = \sum_{i=1}^{D^2} \lambda_{i} \rket{r_{i}}\rbra{l_{i}}$, with $\rket{r_{i}}$ and $\rbra{l_{i}}$ respectively the right and left eigenvectors corresponding to the eigenvalue $\lambda_{i}$, then Eq.~\eqref{eq:gsortho} becomes 
 \begin{align}\label{eq:gsortho2}
&\Braket{\Phi[\overline{V},\overline{W},\overline{Y}] | \Psi(Q,R,B)}  = \sum_{\ic\neq k=1}^{D^{2}} \frac{\ec^{\lambda_{i}\LL} - \ec^{\lambda_{k}\LL}}{\lambda_{i}-\lambda_{k}}  \times \nonumber \\ 
& \Tr \Big[ B\otimes\overline{B}  \rket{r_{i}}\rbra{l_{i}} \left(\mathbb{I}\otimes\overline{V} + R\otimes\overline{W} \right)\rket{r_{k}}\rbra{l_{k}} \Big]  \nonumber \\ 
&+\LL\sum_{i=1}^{D^{2}}\ec^{\lambda_{i}\LL} \Tr  \left[ B\otimes\overline{B} \rket{r_{i}}\rbra{l_{i}} \Big(\mathbb{I}\otimes\overline{V} + R\otimes\overline{W} \Big)\rket{r_{i}}\rbra{l_{i}} \right] \nonumber \\
&+ \sum_{i=1}^{D^2} \ec^{\lambda_{i}\LL} \Tr \Big[ B\otimes\overline{Y} \rket{r_{i}}\rbra{l_{i}}\Big] \; ,
 \end{align} 
where we neglected (accidental) degeneracies of the sub-dominant $\lambda$'s (but a generalization thereto is straightforward).
The vector $[y_{1};y_{2};y_{3}]$ is then immediately found by reshaping vectors to matrices according to  the isomorphism $a\otimes\overline{b}\rket{x}\to a x b^{\dagger}$  (with $a,b,x \in\mathbb{C}^{D\times D}$) s.t. Eq.\eqref{eq:gsortho2} equals $\left[V^{\dagger} W^{\dagger} Y^{\dagger}\right]  \left[ y_{1};y_{2};y_{3}\right]$. However, in order to have an overall computational cost of $\order (D^{3})$ we can not compute the whole spectrum of $T$ as the cost of this would be of $\order (D^{6})$. The situation can be  remedied by introducing a cutoff in the spectrum of $T$ and only calculating the leading $m_{1}$ eigenvalues. The cutoff is chosen in such a way that $\forall i > m_{1}$ ,  $|\ec^{\LL\lambda_{i}}| < tol$ for some predefined tolerance $tol$ and we can set $\ec^{\LL\lambda_{i}} =0$.  For large enough $\LL$ it can happen that the factor $\LL\ec^{\LL\lambda_{i}}$ might again be above the tolerance and hence another cutoff at $m_{2}\geq m_{1}$ needs to be taken into account.  As long as  $m_{2}$ is not  too large this can be done with an iterative eigenvalue solver at a cost of $\order (D^{3})$. The tolerance was chosen to be $tol = 10^{-9}$ in all simulations which roughly corresponds to $m_{2} \approx 30$ for the largest bond dimension used in this paper. The final expression of Eq.~\ref{eq:gsortho} with these approximations can be found in Appendix \ref{sec:AppA}. 
It is however important to mention that this expression will contain the matrix $\big(\tilde{T}_{i}^{m_{1}}\big)^{{\rm{p}}} := \Big[ \lambda_{i} \big(\mathbb{I} - \sum_{k=1}^{m_{1}}\rket{r_{k}}\rbra{l_{k}} \big) - \big( T - \sum_{k=1}^{m_{1}} \lambda_{k}\rket{r_{k}}\rbra{l_{k}} \big) \Big]^{{\rm{p}}}$ where the notation $\rm{p}$ stands for the pseudo-inverse of  a matrix with non vanishing null space (see Appendix (\ref{sec:AppA}) for more details).  At first glance it might seem that we still have to compute the whole spectrum of $T$ in order to calculate the pseudo-inverse of the matrix $\tilde{T}_{i}^{m_{1}}$. This however is not true since we only need to know how $\tilde{T}_{i}^{m_{1}}$ acts on a vector, for which it is sufficient to know only the first $m_{1}$ eigenvalues and eigenvectors. 
This allows us to use an iterative solver to compute the action of $\tilde{T}_{i}^{m_{1}}$ on a vector with a cost of $\order (D^{3})$. The calculation of the Gram matrix $G$ and the vector $[g_{1};g_{2};g_{3}]$ in Eq.~\eqref{eq:TDVPeq} follows along the same lines but is very lengthy and  the explicit calculation can also be found in Appendix (\ref{sec:AppA}). Note that we actually never have to calculate the full matrix $G$ but again only need to know how it acts on a vector. This can also be done with an iterative solver which means that the overall cost of the algorithm remains of $\order (D^{3}) $.

\subsection{The model}\label{sec:secModel}
 
The system we aim to study consists of an interacting Bose gas loaded in a ring shaped trap of length $\rm{L}$ which is interrupted by  a localised barrier rotating at constant velocity.  The importance of this model stems from the fact that its low energy physics allows for the construction of current based quantum information devices such as atomtronic SQUID analogues or flux qubits. In the rotating frame the Hamiltonian describing this system reads 
\begin{align}\label{eq:hamil1}
\operator{H} = \int_{0}^{\LL} \d x \left[ \left(\partial_{x}\operator{\psi}^{\dagger} + \ic\frac{2\pi\Omega}{\LL}\operator{\psi}^{\dagger}\right)\left(\partial_{x}\operator{\psi} - \ic\frac{2\pi\Omega}{\LL}\operator{\psi}\right)  \right. \nonumber \\ 
\left. + c\operator{\psi}^{\dagger}\operator{\psi}^{\dagger}\operator{\psi}\operator{\psi} + U_{0} \delta(x)\operator{\psi}^{\dagger}\operator{\psi}  - \mu \operator{\psi}^{\dagger}\operator{\psi} \right]
\end{align}
with $c > 0$ being the interaction strength, $U_{0}$ the barrier strength, $\mu > 0$ the chemical potential and $\Omega$ the Coriolis flux. The spatial dependence of the operators $\operator{\psi}(x)$ has been omitted for the sake of brevity and we have set $\hbar=1$ and the mass $m=1/2$.  The introduction of a chemical potential is necessary as the cMPS wavefunction in Eq. \eqref{eq:cMPSfunc} is formulated in the grand-canonical picture and does not allow for a fixed number of particles. However, all simulations in section \ref{sec:secResults}  were performed at a fixed average number of particels $N$ for which the chemical potential had to be fine tuned accordingly. An essential feature of the ground state physics of $\operator{H}$ is the periodicity of its ground state energy as a function of $\Omega$ with period $1$ for all values of $c$ and $U_{0}$. This can be appreciated by performing the canonical transformation $\operator{\psi}(x)\to \ec^{\ic \frac{2\pi n}{L} x}\operator{\psi}(x)$ which shifts $\Omega \to \Omega - n$ and preserves the periodic boundary conditions $\psi(0) = \psi(L)$ for integer values $n$. More generally, by using a non-integer value $n=\Omega$, the model is mapped to a standard Lieb-Liniger type Hamiltonian \cite{PhysRev.130.1605}, with the addition of a barrier where the field operators also obey twisted boundary conditions $\operator{\psi}(L) = \ec^{\ic 2\pi \Omega} \operator{\psi}(0)$. At this point it seems that there are two strategies to proceed: one can either simulate the Hamiltonian in \eqref{eq:hamil1} where the field operators obey periodic boundary conditions or one can study the Lieb-Liniger type Hamiltonian with twisted boundary conditions. In both cases the cMPS wave function has to obey a certain regularity condition \cite{PhysRevB.88.085118} to have a well defined kinetic energy at position $x=0$. For a system with pbc this regularity condition reads $BR - RB = 0$ whereas for twisted boundary conditions it reads  $BR - \ec^{ \ic2\pi\Omega}RB = 0$. Both of these conditions impose a certain parametrisation on $R$ and $B$. We have observed  that either parametrisation of $R$ and $B$ negatively affects the numerical stability of the Gram matrix G, even for small values of the bond dimension $D$. 

Therefore, we have chosen a third strategy, which produces a regularised Hamiltonian $\operator{H}_{\epsilon}$ in terms a cut-off length $\epsilon$ \emph{only} around the barrier, where the twisted boundary conditions have not to be enforced explicitly but rather emerge from a (divergent) boundary term in the Hamiltonian.
In order to construct it, we pass through a discretization on a lattice with $N_{\rm{s}} = L/\epsilon$ sites and lattice spacing $\epsilon$, where each site $j$ accommodates modes 
$\tilde{a}_{j} = \sqrt{\epsilon}\, \ec^{ \ic\frac{2\pi\Omega}{\LL}(j - 1/2)\epsilon} \,  \hat{\psi}\left((j - 1/2) \epsilon\right)$.
This is very similar to the above mentioned lattice version commonly used for standard MPS (see e.g. Ref.~\cite{PhysRevLett.113.025301}), apart from a slightly more convenient choice about the barrier delta function, $\delta(x) = \lim_{\epsilon \to 0} \left( \delta(x-\epsilon/2) + \delta(x+\epsilon/2) \right)/2$.
We expand the Hamiltonian~\eqref{eq:hamil1} to the lowest relevant order in $\epsilon$ as
\begin{align}\label{eq:hamillattice}
\hat{H}_{\rm{Lattice}} = &  
\sum_{j=1}^{N_\mathrm{s} -1} \frac{1}{\epsilon^{2}} \big[ \tilde{a}^{\dagger}_{j+1}- \tilde{a}^{\dagger}_{j} \big] \big[ \tilde{a}_{j+1}- \tilde{a}_{j} \big] \nonumber \\
&+\frac{1}{\epsilon^{2}} \big[ \tilde{a}^{\dagger}_{1}- \ec^{ -\ic2\pi\Omega}\tilde{a}^{\dagger}_{N_\mathrm{s}} \big] \big[ \tilde{a}^{}_{1}- \ec^{ \ic2\pi\Omega}\tilde{a}^{}_{N_\mathrm{s}} \big] \nonumber \\
& + \sum_{j=1}^{N_\mathrm{s}}  
\left[ \frac{c}{\epsilon} \tilde{a}^{\dagger}_{j+1} \tilde{a}^{\dagger}_{j} \tilde{a}^{}_{j} \tilde{a}^{}_{j+1}
- \mu \tilde{a}^{\dagger}_{j} \tilde{a}^{}_{j} \right]  
\nonumber \\
& + \frac{U_{0}}{2\epsilon}\big[ \tilde{a}^{\dagger}_{1}\tilde{a}_{1} + \tilde{a}^{\dagger}_{N_\mathrm{s}}\tilde{a}_{N_\mathrm{s}} \big]  \;.
\end{align}
We notice that all complex phases have cancelled in the bulk of the system and $\Omega$ only appears in the hopping across the boundary, thus making the periodicity 1 in $\Omega$ manifest. By taking again the continuum limit and reintroducing the field derivatives, we get

\begin{widetext}
\begin{align}\label{eq:hamileps}
\operator{H}_{\epsilon} = &\int_{0}^{\LL} \d x \; \left[ 
\partial_{x}\operator{\psi}^{\dagger} \partial_{x}\operator{\psi} 
+  c\operator{\psi}^{\dagger}\operator{\psi}^{\dagger}\operator{\psi}\operator{\psi} - \mu \operator{\psi}^{\dagger}\operator{\psi} 
\right]
+ \frac{U_{0}}{2} \big(\operator{\psi}^{\dagger}(0)\operator{\psi}(0)  + \operator{\psi}^{\dagger}(L)\operator{\psi}(L) \big)  \nonumber \\
&+\frac{1}{\epsilon}\big( \operator{\psi}^{\dagger}(0) - \ec^{ -\ic2\pi\Omega}\operator{\psi}^{\dagger}(L)\big) \big( \operator{\psi}(0) - \ec^{ \ic2\pi\Omega}\operator{\psi}(L)\big) \nonumber \\
 &+\frac{1}{2} \big( \operator{\psi}^{\dagger}(0) - \ec^{ -\ic2\pi\Omega}\operator{\psi}^{\dagger}(L)\big) \big( \partial_{x}\operator{\psi}(0) + \ec^{ \ic2\pi\Omega}\partial_{x}\operator{\psi}(L)\big)  +\frac{1}{2} \big( \partial_{x}\operator{\psi}^{\dagger}(0) + \ec^{ -\ic2\pi\Omega}\partial_{x}\operator{\psi}^{\dagger}(L)\big) \big( \operator{\psi}(0) - \ec^{ \ic2\pi\Omega}\operator{\psi}(L)\big) \nonumber\\
 &+ \mathcal{O}(\epsilon)\;.
\end{align}
\end{widetext}
While in this regularized Hamiltonian $\operator{H}_{\epsilon}$ the field operators in Eq.~\eqref{eq:hamileps} are assumed to obey pbc but
the matrices $R$ and $B$ are now completely free, and we can thus circumvent the Gram matrix instability.
This introduces a competition between minimizing the first line in \eqref{eq:hamileps} thus the energy and minimizing the remaining two lines which try to enforce the boundary condition, via a diverging weight.
A detailed study of the energy scaling w.r.t $\epsilon$ is given in Appendix (\ref{sec:AppB}) and unless otherwise stated we will set $\epsilon = 2.5\times1\ec^{-3}$ for the remaining part of the paper. 

\subsection{Results}\label{sec:secResults}

\begin{figure}
\includegraphics[width=43mm,height=45mm]{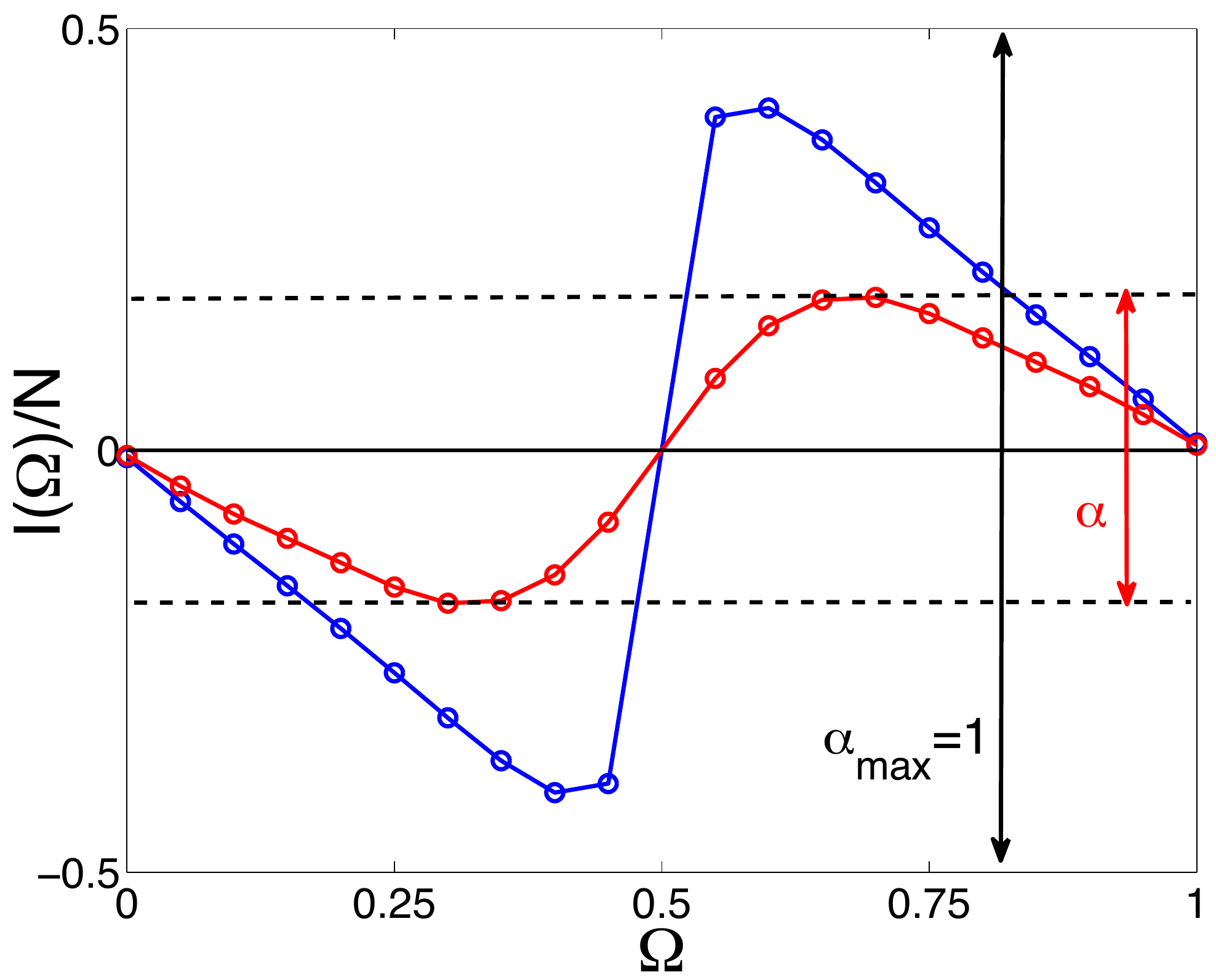}
\includegraphics[width=42mm,height=45mm]{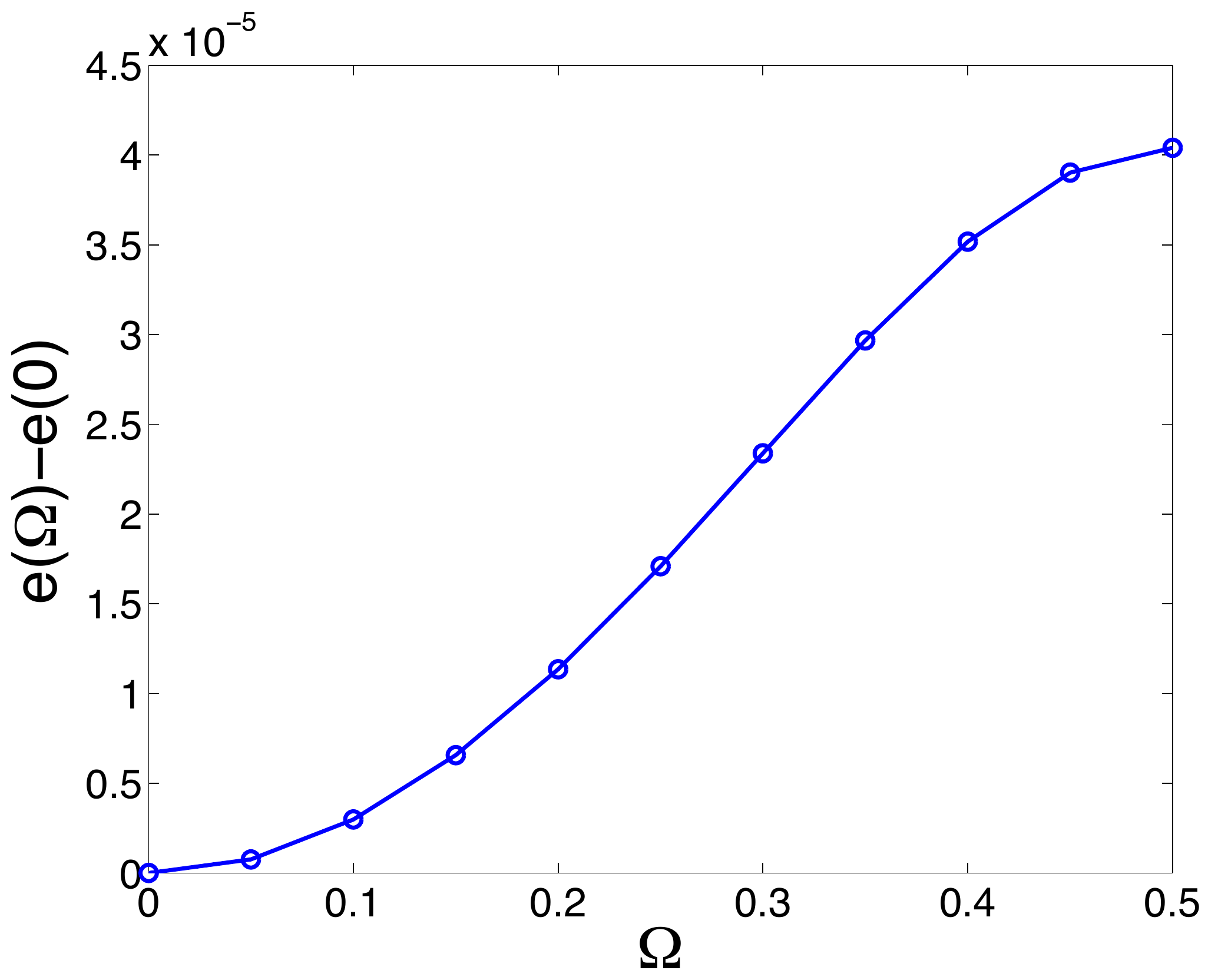}
\caption{\small{Left: Persistent current in units of $I_{0}=2\pi/mL^{2}$ for $\lambda=9.5$ (blue, D=10) and $\lambda=38.2$ (red, D=10) both at $\gamma=2.33$. Right: Energy density  difference at $\lambda = 38.2$ and $\gamma=33$ at $D=12$. The other parameters in both panels are $L=120$ and $N=18$. }}
\label{fig:currentplot}
\end{figure}

In the absence of a barrier the system is a perfect superfluid and its ground state energy is given by a series of intersecting parabolas due to Galiliean invariance  \cite{PhysRevA.7.2187}. The corresponding persistent current $I(\Omega)$ is thus a perfect sawtooth in the rotating frame and it is related to the ground state energy $E(\Omega)$ via
\begin{equation}\label{eq:current}
I(\Omega) = -\frac{1}{2\pi}\frac{\partial E(\Omega)}{\partial\Omega} \;.
\end{equation}   
If the barrier is turned on rotational symmetry is broken and the persistent current changes its shape from a smeared sawtooth (weak barrier) to a sinusoid (large barrier) (see Fig.~\ref{fig:currentplot}). The rich interplay between quantum fluctuations, interactions and statistics manifests itself in the intriguing behaviour of the persistent currents amplitude $\alpha$.  The dependence of the amplitude on both the dimensionless interaction strength $\gamma = c/\rho$ ($\rho$ being the ground state density) and on the barrier height  $\lambda = mU_{0}L/\pi$  was recently thoroughly studied  by some of us~\cite{PhysRevLett.113.025301}.  There the authors combined both analytical (Gross-Pitaevskii and Luttinger-Liquid theory) and numerical (DMRG) techniques to show that the amplitude displays a maximum at intermediate interactions for all values of $\lambda$. In the following we will give numerical evidence that the cMPS not only  shows perfect agreement of the density profiles  when compared to classical GP and DMRG results for the whole parameter range of $\gamma$ and $\lambda$ but also perfectly well captures the non monotonicity of the function $\alpha(\gamma)$. The numerical results are compared to DMRG based lattice simulations and to exact results from a classical Gross-Pitaevskii analysis at weak interactions.

   \emph{Density profiles---}
 \begin{figure*}
\includegraphics[width=80mm,height=40mm]{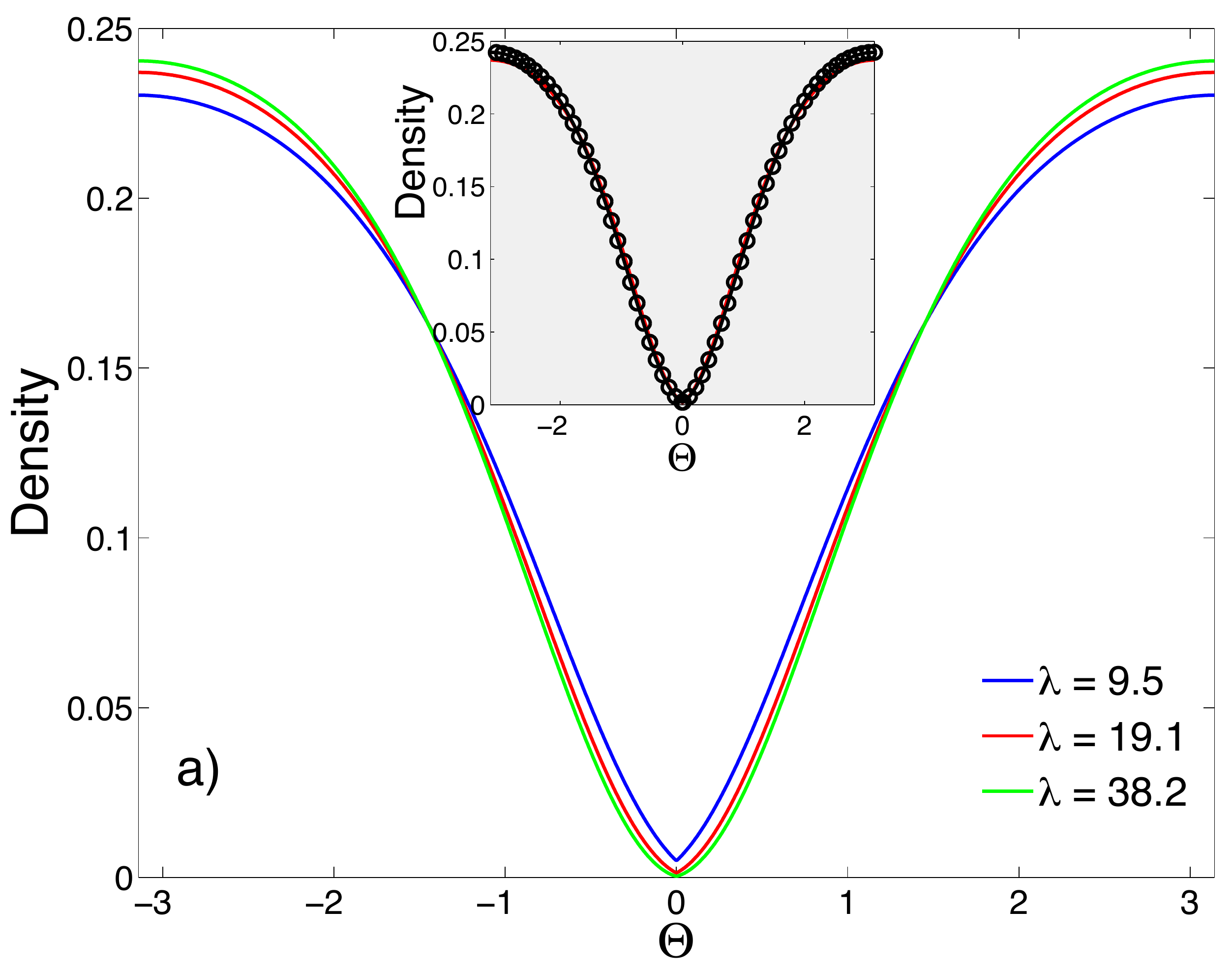}
\includegraphics[width=80mm,height=40mm]{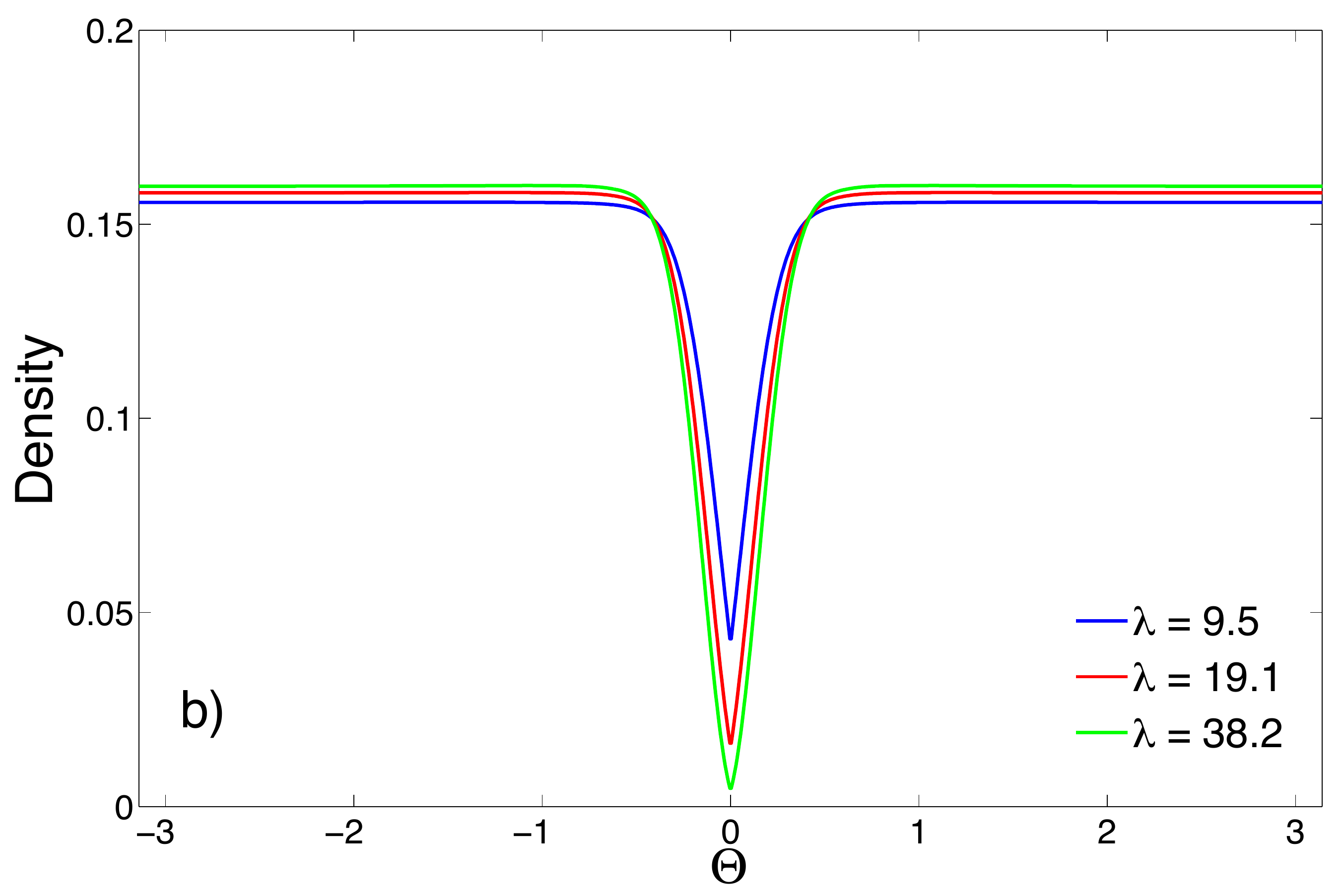}
\includegraphics[width=80mm,height=40mm]{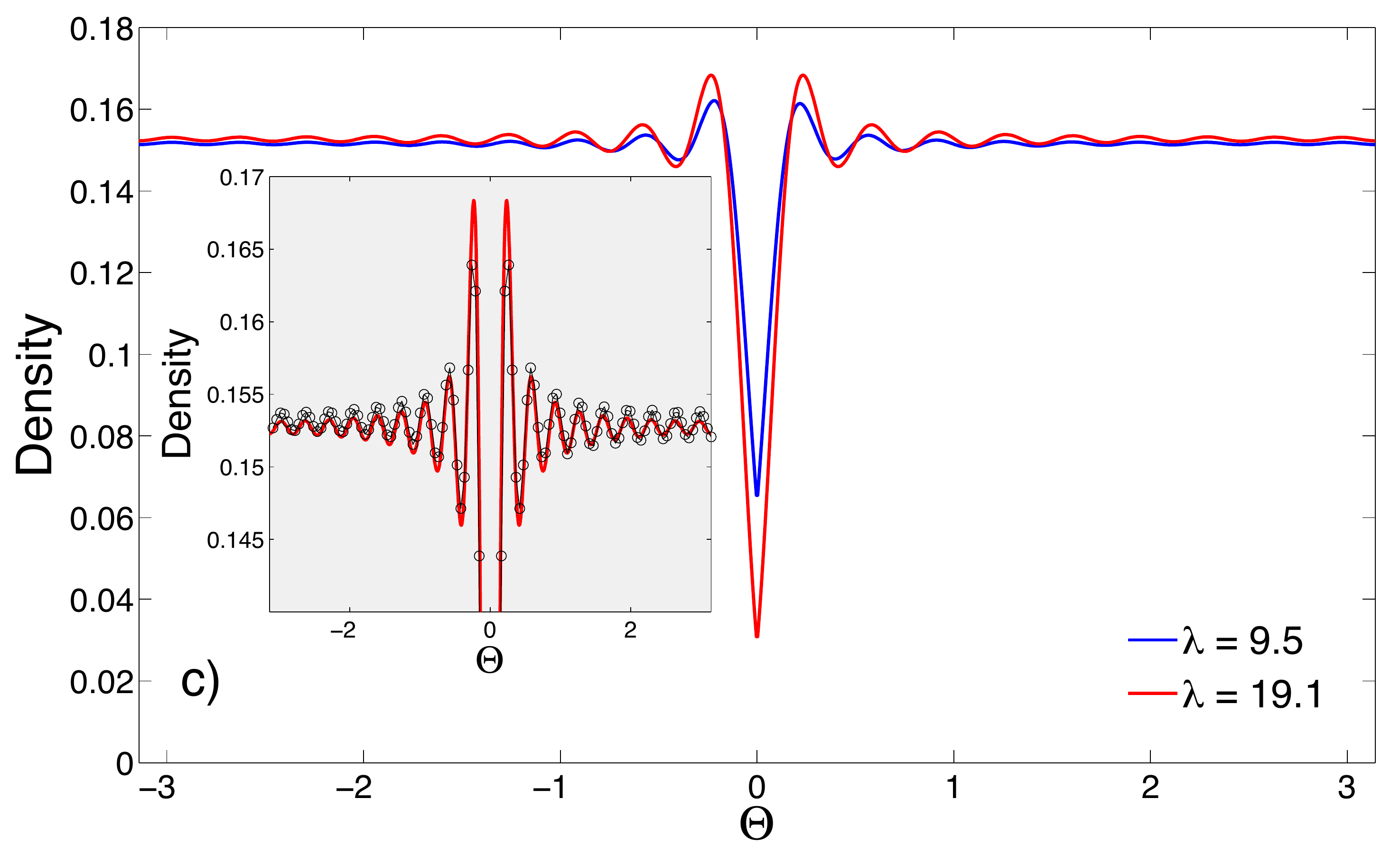}
\includegraphics[width=80mm,height=40mm]{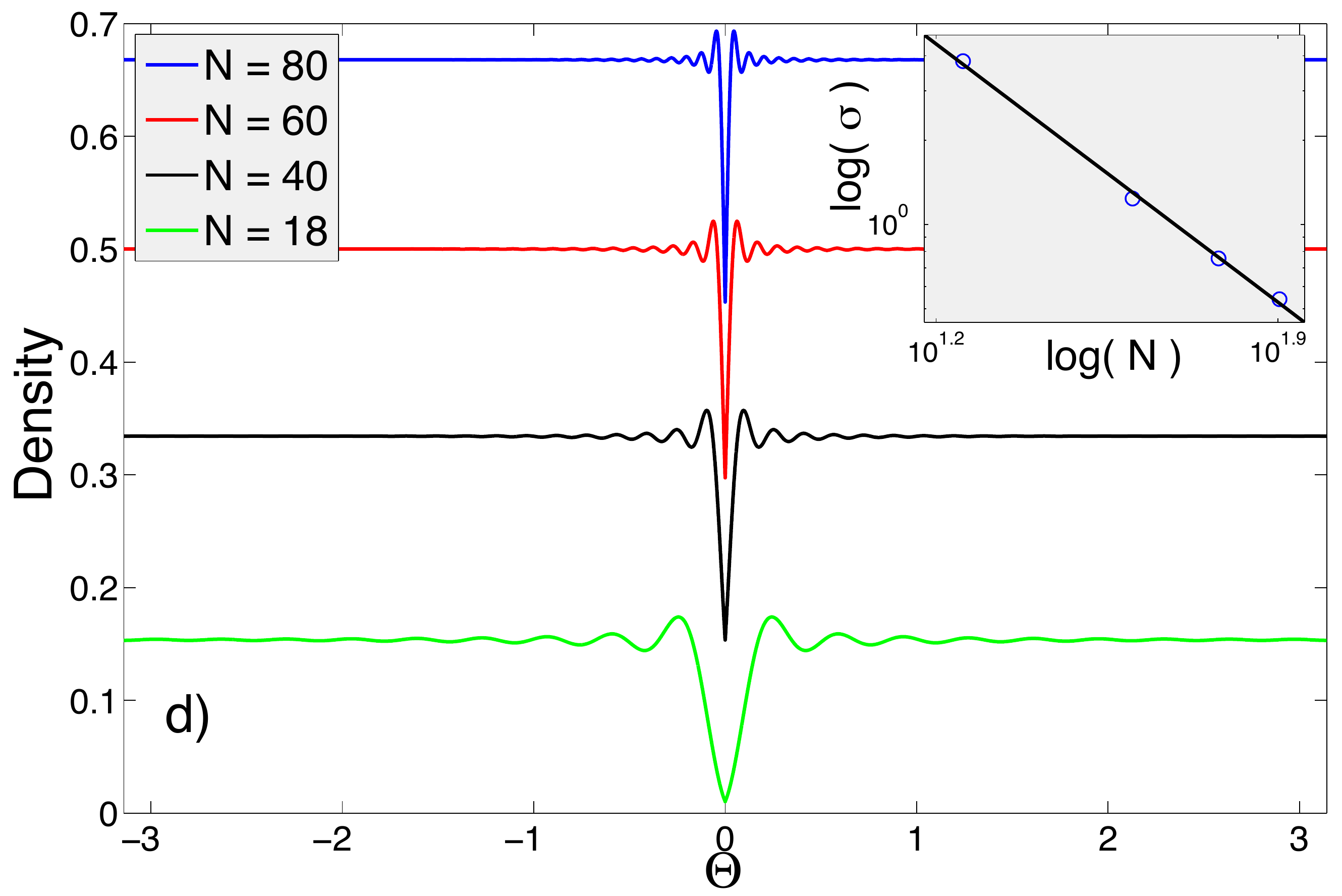}
\caption{\small{Density profiles vs angular coordinate $\Theta = \frac{2\pi}{L}{x}$ at $L=120$ and $\Omega=0$ for different values of the interaction $\gamma$ and the barrier height $\lambda$. (a) Weak interaction at $\gamma=0.03$, $N=18$ and $D=4$. Inset: GP solution (black dots) and cMPS result (red) at $\lambda=19.1$  (b) Intermediate interaction at $\gamma=2.33$, $ N=18$ and $D=10$. (c) Strong interaction at $\gamma=33$, $N=18$ and $D=12$. Inset: DMRG data (black dots) compared to cMPS results (red) at $\lambda=19.1$ . (d) Density profile at fixed $\gamma=33$, $\lambda=19.1$ and $L=120$ but for different expectation value of the particle number $N=80 (D=18), N=60 (D=18), N=40 (D=16)$ and $N=18 (D=12)$. Inset: loglog plot of the width $\sigma$ (blue dots) of the central depletion around $\Theta=0$ as a function of $N$. The black line is a linear fit indicating an algebraic decay of $\sigma(N)=A N^{\eta}$ with $A$ a constant and the exponent $\eta \approx -1.3$.
}}
\label{fig:densityplots}
\end{figure*}
In Fig.~\ref{fig:densityplots} several density profiles are shown for different values of $\gamma$ and barrier height $\lambda$. At small $\gamma$ little entanglement is present in the system and the ground state is well described by a dark soliton solution of the classical Gross-Pitaevskii (GP) equation \cite{PhysRevLett.113.025301}. Hence, the cMPS ansatz already converged to this  solution at a very small bond dimension of $D=4$ as can be seen from the inset of panel (a) in Fig.~\ref{fig:densityplots}. The condition number of the Gram matrix in Eq.~\eqref{eq:TDVPeq} is in general negatively affected by small eigenvalues of the fix point of the transfer matrix $T$. This fact in practice upper bounds the largest possible bond dimension which explains why only rather small values of $D$ could be reached in the limit of weak interaction or little entanglement. With increasing  $\gamma$ the width of the central depletion around $x=0$ decreases as the barrier effectively gets screened by a decaying healing length $\xi = 1/\sqrt{4mc\rho}$. At large $\gamma$ Friedel oscillations start to build up as the density profiles and the energy of the system start to resemble those of free fermions. The inset of panel (c) in Fig.~\ref{fig:densityplots} compares the cMPS result for large $\gamma$ with the  DMRG data from Ref.~\onlinecite{PhysRevLett.113.025301} and shows perfect agreement of the density profiles even at moderate values of $D$. 

The cMPS ansatz also allows to study in principle arbitrary large (or small) densities where it is not possible anymore to describe the system with a low filling lattice approximation. This methodological advantage can be exploited to study e.g. the full width at half maximum of the central depletion, a quantity which could be measured in experiments akin those in Ref.~\onlinecite{HysteresisCamp}. We find that the width is decaying algebraically with the average number of particles when all other parameters are kept fixed as is shown in panel (d) of Fig.~\ref{fig:densityplots}. Note that the density profiles in general converge from inside to the outside as the effective cMPS correlation length increases with increasing $D$.  In the plots of panel (d) the effective correlation length has not fully converged as the Friedel oscillations are too strongly suppressed when compared to results from Luttinger liquid theory \cite{0953-4075-37-7-051}.
 However, the profile of the central depletion and its vicinity and thus the width have already converged which was further checked by a finite $D$ scaling as is shown in Appendix (\ref{sec:AppB}) of this paper.

\emph{Persistent current amplitude---}
 \begin{figure}
\includegraphics[width=85mm,height=55mm]{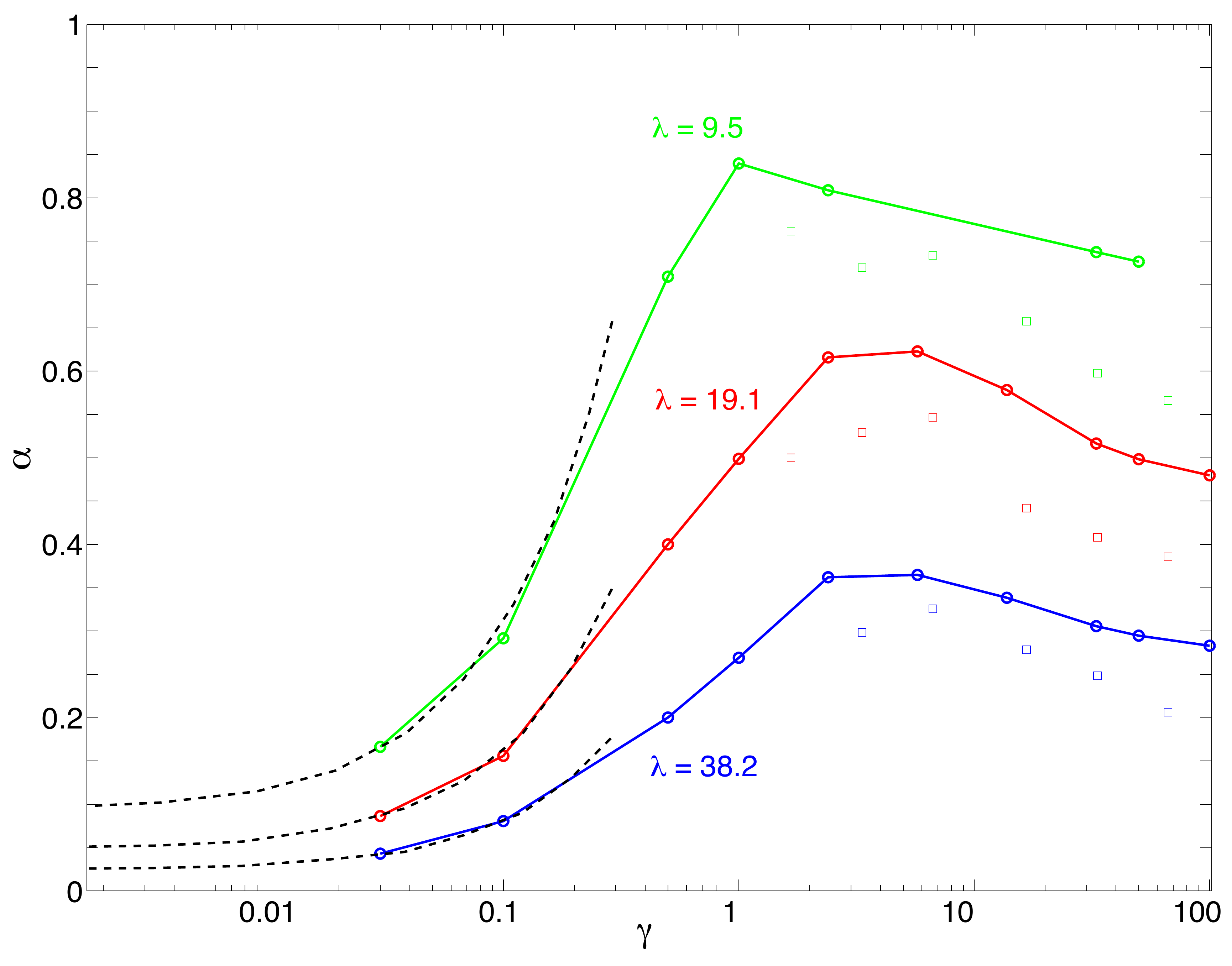}
\caption{\small{Persistent current amplitude in units of $I_{0}=2\pi/mL^{2}$ as a function of $\gamma$ for several values of $\lambda$ at  $L=120$ and $N=18$. Three different approaches were used: cMPS (dots), DMRG (squares) at $D_{\rm{MPS}}=20$ (and unitary lattice spacing $a=1$) and GP (black dashed line). The bond dimension increases from $D=4$ at $\gamma=0.03$ up to $D=14$  for $\gamma\geq 5$.}}
\label{fig:amplitudeplot}
\end{figure}
We will now study the  behaviour of the persistent current amplitude as a function of  $\lambda$ and $\gamma$. The results presented in Fig.~\ref{fig:amplitudeplot} are compared to a DMRG based MPS simulation for intermediate and large $\gamma$ and a classical GP calculation at small $\gamma$. It is worth mentioning that the DMRG based lattice approximation is not applicable in the limit of small $\gamma$ or large densities. In this limit increasing quantum density fluctuations demand for a large occupation number of the local Hilbert basis on the lattice thus making the DMRG simulation computationally inefficient. Moreover, the cosine form of the single particle dispersion relation on the lattice is only at small densities a reasonable  approximation of the parabolic dispersion in the continuum. The cMPS itself however is per definition a superposition of states with different particle number and as thus captures the correct physics even at small $\gamma$. In fact, the cMPS approach perfectly well describes the non monotonic behaviour of $\alpha$ for the whole range of $\gamma$ and $\lambda$. 
With increasing $\gamma$ the bond dimension $D$ has to be increased, since the state gets more and more correlated and sensibly departs from the nearly product state at $\gamma \simeq 0$. The fixed point of $T$ has a fairly large spectral support, and at $\gamma \gg 1$ the largest possible value of $D$ ($D=20$ in Fig.~\ref{fig:TDplot}) is dictated rather by the computational resources at hand than by the condition number of $G$. The cMPS ansatz hence  interpolates between the classical mean-field solution at small $\gamma$ and the strongly correlated quantum regime at large $\gamma$. This beautifully  illustrates the fact that the TDVP equations for cMPS do collapse to the classical GP-equation at $D=1$ which was recently shown in Ref.~\onlinecite{QGP}.

\emph{Thermodynamic limit---}
We are now interested in approaching the thermodynamic limit s.t. both $N,L \rightarrow \infty$ at fixed density $\rho = N/L$. It is well known \cite{PhysRev.130.1605} that the ground state energy of the Lieb-Liniger Hamiltonian scales as $E^{\rm{LL}}_{0} = N \rho^{2} e(\gamma) = L \rho^{3} e(\gamma)$ with $e$ being a dimensionless function depending only on $\gamma$. In the presence of a barrier such a scaling is a bit cumbersome as the barrier height $\lambda = mUL/\pi$ itself also depends on $L$. We will therefore propose a different definition of the barrier height given by
\begin{equation}\label{eq:barrierheight}
\Lambda \equiv m U_{0} \frac{L}{N} = \frac{\pi \lambda}{N} \;.
\end{equation}  
This definition also holds in the thermodynamic limit and it is easy to see that under the scaling transformation $x\rightarrow 1/\rho x$ the ground state energy of  \eqref{eq:hamil1} scales as $E = N \rho^{2} f(\gamma,\Lambda)$ with $f$ being now dimensionless as well. Note that the energy dependence on $\Omega$ has dropped since the corresponding contributions are only of  $\mathcal{O}(1/L)$. In Fig.~\ref{fig:TDplot} the thermodynamic limit scaling of the ground state energy density is shown at $\gamma=80, \Lambda=4 $ and $\rho=0.125$. The cMPS results are further compared to a Tree-Tensor Network (TTN) \cite{PhysRevB.90.125154} based lattice simulation. The TTN approach is based on a lattice discretization of the Hamiltonian given in Eq.~\ref{eq:hamil1}, just like the one used for the MPS/DMRG simulations reported above, while the tensor-network structure is taken to be hierarchical: different length-scales in the system are accounted for by different layers of renormalizing tensors.
The loop-free topology of this network allows to drastically reduce the computational cost of pbc simulations by exploiting the internal gauge degrees of freedom. It is thus possible to work with much larger bond dimensions similar to those of obc simulations.  We find that both cMPS and TTN give the same result for the asymptotic value of the energy density up to an error due to the absence of a finite $D$ scaling. Note that all  TTN results were calculated in the  continuum limit by considering  $\lim_{a\rightarrow0}E(a)$ with $a$ being the lattice spacing in the TTN simulation. It is further instructive to compare the number of variational parameters used in both approaches.  In a TTN simulation the number of variational parameters is  $(N_{\rm{sites}}(d-1)D^{2}_{\rm{TTN}})$ where $d$ is the local Hilbert space dimension (bosonic occupation $0 \ldots (d-1)$) in the TTN calculation ($d\approx 4$ in all simulations). In comparison,  the number of variational parameters in the cMPS is only $(2D^{2})$ which is several orders of magnitude smaller compared to the TTN approach for the bond dimensions and system sizes used in Fig.~\ref{fig:TDplot}. To give an example we calculate the energy density at $L=256$,  $\Omega=0$,  $\rho=0.125$, $\gamma=80$ and $\Lambda=4$. The cMPS result at $D=16$ equal $E/L = 0.0063305 (\pm 1\mathrm{e}{-7})$ while the TTN at $D_{\rm{TTN}} = 96$  gives $E/L = 0.00632 (\pm 1\mathrm{e}{-5})$ (with the cMPS error bar coming from a finite $\epsilon > 0$ as explained in Appendix (\ref{sec:AppB}) and the error in the TTN energy stemming from taking the continuum limit).
 Quite remarkably, both energies are well compatible even though in the former case only $512$ variational parameters had to be optimized while for the latter as much as 28 millions (at $d=4$ and $a=1/4$) were taken into account. 
 This a clear advantage of the cMPS ansatz itself independent of the possible variational optimization procedures at hand. However, in practice the efficiency of an algorithm not only depends on the number of variational parameters or the overall computational scaling of the algorithm but also how many iterations it takes to converge to a desired precision. While a direct performance comparison between cMPS, TTN and DMRG would go beyond the intention of this paper we do however want to stress that cMPS only compares well if a conjugate gradient optimization (as explained in section \ref{sec:secTDVP}) is applied.      

 \begin{figure}
\includegraphics[width=85mm,height=55mm]{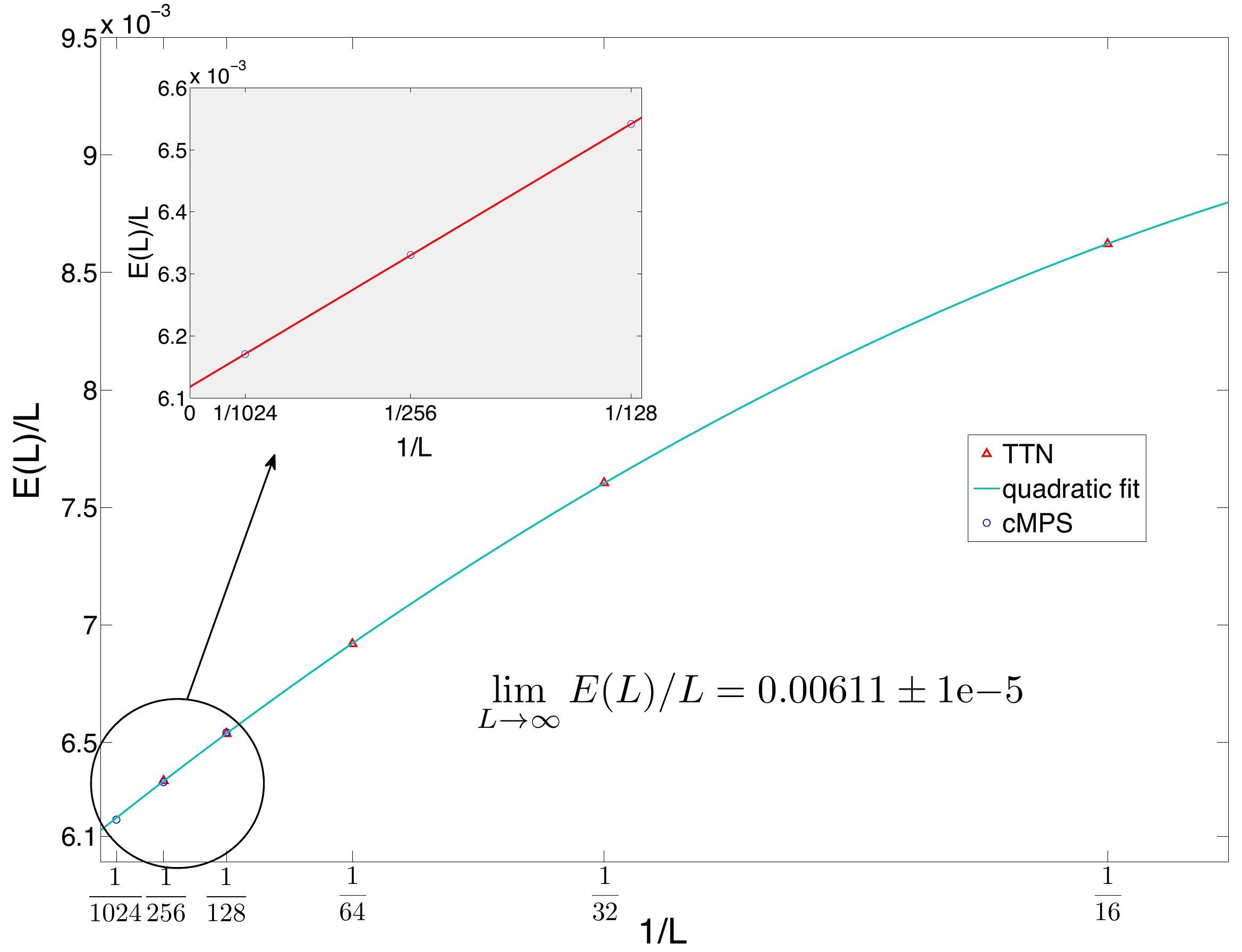}
\caption{\small{Energy density for different system sizes. The TTN simulation (red triangles) was performed at $D_{\rm{TTN}}=96$ while the largest value for the cMPS simulations (blue circles) was $D=20$ at $L=1024$. Inset: Linear fit of the cMPS data. All simulations were done at $\Omega=0$,  $\rho=0.125$, $\gamma=80$ and $\Lambda=4$.  
}}
\label{fig:TDplot}
\end{figure}

\subsection{Outlook and Conclusion}\label{sec:secCon}
   
 We have developed a time evolution algorithm for uniform cMPS with periodic boundary conditions. The translational invariance of the ansatz is slightly broken as the algorithm also allows for local degrees of freedom encoded in a cMPS boundary matrix. The method is then used to study a rotating bose gas at all interactions which is interrupted by a localised barrier. The results are carefully benchmarked against lattice simulations and classical GP calculations showing very good agreement for all interactions.  The study of real-time evolution would have gone beyond the scope of this paper but the algorithm is perfectly well suited for dynamical problems as well. Indeed, it is in principle not only possible to study the dynamics of global quenches but also the ones of local quenches in the barrier strength. Moreover, it is possible to generalize the method to the case of multiple species or fermions to study e.g. spin currents or spinor condensates which have been recently experimentally realised \cite{PhysRevLett.110.025301}.  Note that in both of these cases additional regularity constraints have to be imposed on the $R_{\alpha}$ matrices where the index labels the individual species. In the case of fermions this condition reads $R_{\alpha}^{2}=0$ and a general  parametrization was recently propsed  in \cite{PhysRevLett.105.251601,PhysRevB.91.121108}. It is worth mentioning that the boundary commutation relation with the $B$ matrix would be in general unaffected if the statistics of the field operators changes and thus also the boundary terms in Eq.\eqref{eq:hamileps} would stay the same.  Another direction of future research would be to extend the recently introduced cMPS ansatz for low lying excitations \cite{PhysRevLett.111.020402} to the pbc setting, which would build upon the techniques presented in this paper.

\acknowledgements{
We would like to thank Marco Cominotti for sharing  the GP data used in Fig.~\ref{fig:densityplots} and Fig.~\ref{fig:amplitudeplot} and V. Zauber-Stauner as well as V. Stojevic for inspiring discussions. This project was supported by the Austrian Science Fund (FWF): F4104 SFB ViCoM and F4014 SFB FoQuS, ERA Chemistry and the EC through grants QUTE and SIQS, and by DFG through OSCAR RI2345/2-1.}

\bibliography{bibpbc.bib}

\clearpage
\begin{widetext}
\appendix
\allowdisplaybreaks
 
\section{Calculation of the Gram matrix and the gradient}\label{sec:AppA} 

We start by calculating the overlap of a tangent vector which defines the Gram matrix $G$ via Eq.~(3) in the main text
\begin{align}
\bra{\Phi[\overline{V},\overline{W},\overline{Y}]}\ket{\Phi[V,W,Y]} = &\int_{0}^{\LL}  \d x \; \int_{x}^{\LL}  \d y \; \Tr \Big[ B\otimes\overline{B} \ec^{xT}\Big(V\otimes\mathbb{I} + W\otimes\overline{R} \Big)\ec^{(y-x)T}\Big(\mathbb{I}\otimes\overline{V} + R\otimes\overline{W} \Big)\ec^{(\LL-y)T}  \nonumber \\
& \;\;\;\;\;\;\;\;\;\;\;\;\;\;\;\;\;\;\;\;\;\;\;\; +B\otimes\overline{B} \ec^{xT}\Big(\mathbb{I}\otimes\overline{V} + R\otimes\overline{W} \Big)\ec^{(y-x)T} \Big(V\otimes\mathbb{I} + W\otimes\overline{R} \Big) \ec^{(\LL-y)T} \Big]  \nonumber \\
& +\int_{0}^{\LL}  \d x  \; \Tr \Big[ B\otimes\overline{B} \ec^{xT} W\otimes\overline{W}  \ec^{(\LL-x)T}\Big]   \nonumber \\
& +\int_{0}^{\LL}  \d x  \; \Tr \Big[ B\otimes\overline{Y} \ec^{xT} \Big(V\otimes\mathbb{I} + W\otimes\overline{R} \Big)  \ec^{(\LL-x)T} \Big]  \nonumber \\
& +\int_{0}^{\LL}  \d x  \; \Tr \Big[ Y\otimes\overline{B} \ec^{xT} \Big(\mathbb{I}\otimes\overline{V} + R\otimes\overline{W} \Big)  \ec^{(\LL-x)T} \Big]  + \Tr \Big[ Y\otimes\overline{Y} \ec^{LT}\Big] \;. 
\end{align}
To compute these integrals we insert a spectral decomposition of the transfer matrix $T$ and further define
\begin{equation}\label{eq:generalizedT}
\tilde{T}_{i}^{m_{j}} \Big|_{i=1}^{m_{j}} := \sum_{k=m_{j}+1}^{D^{2}} (\lambda_{i} - \lambda_{k}) \rket{r_{k}}\rbra{l_{k}} =\Big[ \lambda_{i} \big(\mathbb{I} - \sum_{k=1}^{m_{j}}\rket{r_{k}}\rbra{l_{k}} \big) - \big( T - \sum_{k=1}^{m_{j}} \lambda_{k}\rket{r_{k}}\rbra{l_{k}} \big) \Big]
\end{equation}
and $\operator{T}_{i}^{m_{j}} := \left(\tilde{T}_{i}^{m_{j}}\right)^{2}$ where $\rket{r_{k}}\rbra{l_{k}}$ is the projector on the eigenspace with eigenvalue $\lambda_{k}$ of $T$ (we assume that the eigenvalues are non-degenerate) and the integers $m_{j}<D^{2}$ being the cutoff. The pseudo inverse of these matrices is denoted by $\left(\tilde{T}_{i}^{m_{j}}\right)^{\rm{p}} = \sum_{k=m_{j}+1}^{D^{2}} \frac{1}{(\lambda_{i} - \lambda_{k})} \rket{r_{k}}\rbra{l_{k}}$ whose action on a vector can be iteratively computed at a computational cost $\mathcal{O}(D^{3} + m_j D^2)$ by making use of the r.h.s. of Eq.~\ref{eq:generalizedT}:.   We also introduce a third cutoff $m_{3}$ such that $\forall i > m_{3}$, $\LL^{2}/2|\ec^{\LL\lambda_{i}}| < tol$. With this definitions the overlap is then found to be as follows.
\begin{align}
&\bra{\Phi[\overline{V},\overline{W},\overline{Y}]}\ket{\Phi[V,W,Y]} \approx  \\ \nonumber
& \Big\{ \sum_{i\neq j\neq k = 1}^{m_{1}} \frac{(\lambda_{i} - \lambda_{k})(\ec^{\lambda_{i}\LL} - \ec^{\lambda_{j}\LL}) - (\lambda_{i} - \lambda_{j})(\ec^{\lambda_{i}\LL} - \ec^{\lambda_{k}\LL})}{(\lambda_{i} - \lambda_{j})(\lambda_{i} - \lambda_{k})(\lambda_{j} - \lambda_{k})} \nonumber \\
&\qquad\qquad\qquad\qquad\qquad\qquad\qquad\qquad\quad \times\Tr \Big[ B\otimes\overline{B}\rket{r_{i}}\rbra{l_{i}} \Big(V\otimes\mathbb{I} + W\otimes\overline{R} \Big)  \rket{r_{j}}\rbra{l_{j}}   \Big(\mathbb{I}\otimes\overline{V} + R\otimes\overline{W} \Big)  \rket{r_{k}}\rbra{l_{k}} \Big] \nonumber \\ 
&+\left(\sum_{i\neq j = 1}^{m_{2}}  \frac{\LL\ec^{\lambda_{j}\LL}}{(\lambda_{j} - \lambda_{i})} + \sum_{i\neq j = 1}^{m_{1}}  \frac{\ec^{\lambda_{i}\LL} - \ec^{\lambda_{j}\LL}}{(\lambda_{j} - \lambda_{i})^{2}} \right) \Tr \Big[ B\otimes\overline{B}\rket{r_{i}}\rbra{l_{i}} \Big(V\otimes\mathbb{I} + W\otimes\overline{R} \Big)  \rket{r_{j}}\rbra{l_{j}}   \Big(\mathbb{I}\otimes\overline{V} + R\otimes\overline{W} \Big)  \rket{r_{j}}\rbra{l_{j}} \nonumber \\ 
&\qquad\qquad\qquad\qquad\qquad\qquad\qquad\qquad\qquad\;\;+  B\otimes\overline{B}\rket{r_{j}}\rbra{l_{j}} \Big(V\otimes\mathbb{I} + W\otimes\overline{R} \Big)  \rket{r_{i}}\rbra{l_{i}}   \Big(\mathbb{I}\otimes\overline{V} + R\otimes\overline{W} \Big)  \rket{r_{j}}\rbra{l_{j}} \nonumber \\ 
&\qquad\qquad\qquad\qquad\qquad\qquad\qquad\qquad\qquad\;\;+  B\otimes\overline{B}\rket{r_{j}}\rbra{l_{j}} \Big(V\otimes\mathbb{I} + W\otimes\overline{R} \Big)  \rket{r_{j}}\rbra{l_{j}}   \Big(\mathbb{I}\otimes\overline{V} + R\otimes\overline{W} \Big)  \rket{r_{i}}\rbra{l_{i}} \Big]  \nonumber \\ 
&+\frac{\LL^{2}}{2}\sum_{i = 1}^{m3} \ec^{\lambda_{i}\LL} \Tr \Big[ B\otimes\overline{B}\rket{r_{i}}\rbra{l_{i}} \Big(V\otimes\mathbb{I} + W\otimes\overline{R} \Big)  \rket{r_{i}}\rbra{l_{i}}   \Big(\mathbb{I}\otimes\overline{V} + R\otimes\overline{W} \Big)  \rket{r_{i}}\rbra{l_{i}} \Big]  \nonumber \\ 
 \nonumber \\
&+\sum_{i\neq j = 1}^{m_{1}}  \Tr \Big[ B\otimes\overline{B}\rket{r_{i}}\rbra{l_{i}} \Big(V\otimes\mathbb{I} + W\otimes\overline{R} \Big)  \rket{r_{j}}\rbra{l_{j}}   \Big(\mathbb{I}\otimes\overline{V} + R\otimes\overline{W} \Big) \Big( \frac{\ec^{\lambda_{i}\LL}}{(\lambda_{i} - \lambda_{j})}  \Big(\tilde{T}_{i}^{m_{1}}\Big)^{\rm{p}} - \frac{\ec^{\lambda_{j}\LL}}{(\lambda_{i} - \lambda_{j})}  \Big(\tilde{T}_{j}^{m_{1}}\Big)^{\rm{p}} \Big)  \nonumber \\ 
&+\sum_{i\neq k = 1}^{m_{1}}  \Tr \Big[ B\otimes\overline{B}\rket{r_{i}}\rbra{l_{i}} \Big(V\otimes\mathbb{I} + W\otimes\overline{R} \Big)     \Big( \frac{\ec^{\lambda_{i}\LL}}{(\lambda_{i} - \lambda_{k})}  \Big(\tilde{T}_{i}^{m_{1}}\Big)^{\rm{p}} - \frac{\ec^{\lambda_{k}\LL}}{(\lambda_{i} - \lambda_{k})}  \Big(\tilde{T}_{k}^{m_{1}}\Big)^{\rm{p}} \Big)   \Big(\mathbb{I}\otimes\overline{V} + R\otimes\overline{W} \Big) \rket{r_{k}}\rbra{l_{k}} \Big]  \nonumber \\ 
&+\sum_{j\neq k = 1}^{m_{1}}  \Tr \Big[ B\otimes\overline{B} \Big( \frac{\ec^{\lambda_{j}\LL}}{(\lambda_{j} - \lambda_{k})}  \Big(\tilde{T}_{j}^{m_{1}}\Big)^{\rm{p}} - \frac{\ec^{\lambda_{k}\LL}}{(\lambda_{j} - \lambda_{k})}  \Big(\tilde{T}_{k}^{m_{1}}\Big)^{\rm{p}} \Big)  \Big(V\otimes\mathbb{I} + W\otimes\overline{R} \Big)   \rket{r_{j}}\rbra{l_{j}}    \Big(\mathbb{I}\otimes\overline{V} + R\otimes\overline{W} \Big) \rket{r_{k}}\rbra{l_{k}} \Big]  \nonumber \\ 
&  \nonumber \\ 
&+\sum_{i= 1}^{m_{1}} \ec^{\lambda_{i}\LL} \Tr \Big[ B\otimes\overline{B}\rket{r_{i}}\rbra{l_{i}} \Big(V\otimes\mathbb{I} + W\otimes\overline{R} \Big)   \Big(\tilde{T}_{i}^{m_{1}}\Big)^{\rm{p}}   \Big(\mathbb{I}\otimes\overline{V} + R\otimes\overline{W} \Big)  \Big(\tilde{T}_{i}^{m_{1}}\Big)^{\rm{p}} \Big]  \nonumber \\ 
 &+\sum_{j = 1}^{m_{1}} \ec^{\lambda_{j}\LL}  \Tr \Big[ B\otimes\overline{B} \Big(\tilde{T}_{j}^{m_{1}}\Big)^{\rm{p}} \Big(V\otimes\mathbb{I} + W\otimes\overline{R} \Big) \rket{r_{j}}\rbra{l_{j}}    \Big(\mathbb{I}\otimes\overline{V} + R\otimes\overline{W} \Big)  \Big(\tilde{T}_{j}^{m_{1}}\Big)^{\rm{p}} \Big]  \nonumber \\ 
&+\sum_{k = 1}^{m_{1}} \ec^{\lambda_{k}\LL} \Tr \Big[ B\otimes\overline{B} \Big(\tilde{T}_{k}^{m_{1}}\Big)^{\rm{p}} \Big(V\otimes\mathbb{I} + W\otimes\overline{R} \Big) \Big(\tilde{T}_{k}^{m_{1}}\Big)^{\rm{p}}  \Big(\mathbb{I}\otimes\overline{V} + R\otimes\overline{W} \Big)  \rket{r_{k}}\rbra{l_{k}}   \Big]  \nonumber \\ 
&  \nonumber \\ 
&-\sum_{i = 1}^{m_{1}} \ec^{\lambda_{i}\LL} \Tr \Big[ B\otimes\overline{B} \Big(\hat{T}_{i}^{m_{1}}\Big)^{\rm{p}} \Big(V\otimes\mathbb{I} + W\otimes\overline{R} \Big)  \rket{r_{i}}\rbra{l_{i}}  \Big(\mathbb{I}\otimes\overline{V} + R\otimes\overline{W} \Big)  \rket{r_{i}}\rbra{l_{i}}   \Big]  \nonumber \\ 
&-\sum_{i = 1}^{m_{1}} \ec^{\lambda_{i}\LL} \Tr \Big[ B\otimes\overline{B} \rket{r_{i}}\rbra{l_{i}} \Big(V\otimes\mathbb{I} + W\otimes\overline{R} \Big) \Big(\hat{T}_{i}^{m_{1}}\Big)^{\rm{p}}  \Big(\mathbb{I}\otimes\overline{V} + R\otimes\overline{W} \Big)  \rket{r_{i}}\rbra{l_{i}}   \Big]  \nonumber \\ 
&-\sum_{i = 1}^{m_{1}} \ec^{\lambda_{i}\LL} \Tr \Big[ B\otimes\overline{B} \rket{r_{i}}\rbra{l_{i}} \Big(V\otimes\mathbb{I} + W\otimes\overline{R} \Big)  \rket{r_{i}}\rbra{l_{i}} \Big(\mathbb{I}\otimes\overline{V} + R\otimes\overline{W} \Big) \Big(\hat{T}_{i}^{m_{1}}\Big)^{\rm{p}}   \Big]  \nonumber \\ 
&  \nonumber \\ 
&+L\sum_{i = 1}^{m_{2}} \ec^{\lambda_{i}\LL} \Tr \Big[ B\otimes\overline{B} \Big(\tilde{T}_{i}^{m_{2}}\Big)^{\rm{p}} \Big(V\otimes\mathbb{I} + W\otimes\overline{R} \Big)  \rket{r_{i}}\rbra{l_{i}}  \Big(\mathbb{I}\otimes\overline{V} + R\otimes\overline{W} \Big)  \rket{r_{i}}\rbra{l_{i}}   \Big]  \nonumber \\ 
&+L\sum_{i = 1}^{m_{2}} \ec^{\lambda_{i}\LL} \Tr \Big[ B\otimes\overline{B} \rket{r_{i}}\rbra{l_{i}} \Big(V\otimes\mathbb{I} + W\otimes\overline{R} \Big) \Big(\tilde{T}_{i}^{m_{2}}\Big)^{\rm{p}}  \Big(\mathbb{I}\otimes\overline{V} + R\otimes\overline{W} \Big)  \rket{r_{i}}\rbra{l_{i}}   \Big]  \nonumber \\ 
&+L\sum_{i = 1}^{m_{2}} \ec^{\lambda_{i}\LL} \Tr \Big[ B\otimes\overline{B} \rket{r_{i}}\rbra{l_{i}} \Big(V\otimes\mathbb{I} + W\otimes\overline{R} \Big)  \rket{r_{i}}\rbra{l_{i}} \Big(\mathbb{I}\otimes\overline{V} + R\otimes\overline{W} \Big) \Big(\tilde{T}_{i}^{m_{2}}\Big)^{\rm{p}}   \Big]  \nonumber \\ 
& + \Big(V\otimes\mathbb{I} + W\otimes\overline{R} \Big)\leftrightarrow\Big(\mathbb{I}\otimes\overline{V} + R\otimes\overline{W} \Big)  \Big\}
&  \nonumber \\ 
& +   \sum_{i\neq j = 1}^{m_{1}}  \frac{\ec^{\lambda_{i}\LL} - \ec^{\lambda_{j}\LL}}{(\lambda_{i} - \lambda_{j})} \Tr \Big[ B\otimes\overline{B}\rket{r_{i}}\rbra{l_{i}} W\otimes\overline{W}  \rket{r_{j}}\rbra{l_{j}} \Big]  +   \LL \sum_{i=1}^{m_{2}} \ec^{\lambda_{i} \LL}  \Tr \Big[ B\otimes\overline{B}\rket{r_{i}}\rbra{l_{i}} W\otimes\overline{W} \rket{r_{i}}\rbra{l_{i}} \Big] \nonumber \\
& + \sum_{i = 1}^{m_{1}} \ec^{\lambda_{i}\LL}   \Tr \Big[ B\otimes\overline{B} \rket{r_{i}}\rbra{l_{i}} W\otimes\overline{W}  \Big(\tilde{T}_{i}^{m_{1}}\Big)^{\rm{p}} + B\otimes\overline{B} \Big(\tilde{T}_{i}^{m_{1}}\Big)^{\rm{p}}  W\otimes\overline{W} \rket{r_{i}}\rbra{l_{i}} \Big]   \nonumber \\
& +   \sum_{i\neq j = 1}^{m_{1}}  \frac{\ec^{\lambda_{i}\LL} - \ec^{\lambda_{j}\LL}}{(\lambda_{i} - \lambda_{j})} \Big( \Tr \Big[ B\otimes\overline{Y}\rket{r_{i}}\rbra{l_{i}}\Big(V\otimes\mathbb{I} + W\otimes\overline{R} \Big)  \rket{r_{j}}\rbra{l_{j}} \Big]  + \Tr \Big[ Y \otimes\overline{B}\rket{r_{i}}\rbra{l_{i}}  \Big(\mathbb{I}\otimes\overline{V} + R\otimes\overline{W} \Big)   \rket{r_{j}}\rbra{l_{j}} \Big] \Big)  \nonumber \\
& + \sum_{i = 1}^{m_{1}} \ec^{\lambda_{i}\LL}   \Tr \Big[ B\otimes\overline{Y} \rket{r_{i}}\rbra{l_{i}} \Big(V\otimes\mathbb{I} + W\otimes\overline{R} \Big)  \Big(\tilde{T}_{i}^{m_{1}}\Big)^{\rm{p}} + B\otimes\overline{Y} \Big(\tilde{T}_{i}^{m_{1}}\Big)^{\rm{p}}  \Big(V\otimes\mathbb{I} + W\otimes\overline{R} \Big) \rket{r_{i}}\rbra{l_{i}} \Big]   \nonumber \\
& + \sum_{i = 1}^{m_{1}} \ec^{\lambda_{i}\LL}   \Tr \Big[ Y\otimes\overline{B} \rket{r_{i}}\rbra{l_{i}} \Big(\mathbb{I}\otimes\overline{V} + R\otimes\overline{W} \Big) \Big(\tilde{T}_{i}^{m_{1}}\Big)^{\rm{p}} + Y\otimes\overline{B} \Big(\tilde{T}_{i}^{m_{1}}\Big)^{\rm{p}}  \Big(\mathbb{I}\otimes\overline{V} + R\otimes\overline{W} \Big) \rket{r_{i}}\rbra{l_{i}} \Big]   \nonumber \\
& +  \LL \sum_{i=1}^{m_{2}} \ec^{\lambda_{i} \LL} \Big( \Tr \Big[ B\otimes\overline{Y}\rket{r_{i}}\rbra{l_{i}} \Big(V\otimes\mathbb{I} + W\otimes\overline{R} \Big)  \rket{r_{i}}\rbra{l_{i}} \Big]  + \Tr \Big[ Y\otimes\overline{B}\rket{r_{i}}\rbra{l_{i}} \Big(\mathbb{I}\otimes\overline{V} + R\otimes\overline{W} \Big)  \rket{r_{i}}\rbra{l_{i}} \Big]  \Big) \nonumber \\
& + \sum_{i=1}^{m1} \ec^{\lambda_{i} \LL}  \Tr \Big[ Y\otimes\overline{Y}\rket{r_{i}}\rbra{l_{i}} \Big] \;. 
\end{align}
The terms in the above expression are ordered as follows. The curly brackets include those terms where three transfer matrices had to be computed giving rise to threefold indexed expressions. Lines $1-5$ have contributions of all three indices being smaller than the corresponding cutoffs $m_{1}$, $m_{2}$ or $m_{3}$. In the very first line all three indices are unequal  whereas from line 2 to 4 two of them are equal and in the fifth line all three indices are equal to each other. Lines 6-8 are derived from line 1 but where one index exceeds the cut-off. The next block of lines 9-11 contains the sum of terms from line1 where two indices exceed the cutoff plus the terms from lines 2-4 where the index pair which is equal to each other exceeds the cutoff. The last two blocks of lines 12-14 and 15-17 within the curly brackets are derived from lines 2-4 but where the one index which is unequal to the others exceeds the cutoff. The notation in line 18 is an abbreviation for terms which are the same as the ones inside the curly brackets but where $ \Big(V\otimes\mathbb{I} + W\otimes\overline{R} \Big)$ and   $\Big(\mathbb{I}\otimes\overline{V} + R\otimes\overline{W} \Big)$ exchange places. Finally, the last seven lines in the whole expression are terms which include only two transfer matrices and the distinction of cases goes along the same lines.   

We will now proceed by calculating the \textit{gradient} defined in Eq.~(3) of the main text for the Hamiltonian $\operator{H}_{\epsilon}$ defined in the paper.  We again refer to \cite{PhysRevB.88.085118,PhysRevLett.111.020402} for details on cMPS calculus involving a Hamiltonian. The gradient is then defined via the following expression
\begin{align}
&\bra{\Phi[\overline{V},\overline{W},\overline{Y}]}\operator{H}_{\epsilon}\ket{\Psi(Q,R,B)} = \nonumber \\
&\int_{0}^{\LL}  \d x \; \int_{x}^{\LL}  \d y \; \Tr \Big[ B\otimes\overline{B} \ec^{xT} \Big([Q,R]\otimes[\overline{Q},\overline{R}] + c R^{2}\otimes\overline{R}^{2}  - \mu R\otimes\overline{R}\Big) \ec^{(y-x)T}\Big(\mathbb{I}\otimes\overline{V} + R\otimes\overline{W} \Big)\ec^{(\LL-y)T} \nonumber \\ 
& \;\;\;\;\;\;\;\;\;\;\;\;\;\;\;\;\;\;\;\;\;\;\;\; +B\otimes\overline{B} \ec^{xT}\Big(\mathbb{I}\otimes\overline{V} + R\otimes\overline{W} \Big)\ec^{(y-x)T}  \Big([Q,R]\otimes[\overline{Q},\overline{R}] + c R^{2}\otimes\overline{R}^{2}  - \mu R\otimes\overline{R}\Big) \ec^{(\LL-y)T} \Big] \nonumber \\ 
& +\int_{0}^{\LL}  \d x  \; \Tr \Big[ B\otimes\overline{B} \ec^{xT}  \Big([Q,R]\otimes([\overline{Q},\overline{W}] + [\overline{V},\overline{R}]) + cR^{2}\otimes(\overline{R}\overline{W} + \overline{W}\overline{R}) -\mu R\otimes\overline{W} \Big)  \ec^{(\LL-x)T} \Big] \nonumber \\ 
& +\int_{0}^{\LL}  \d x  \; \frac{U_{0}}{2} \Tr \Big[ \Big( BR\otimes\overline{BR} + RB\otimes\overline{RB} \Big) \ec^{xT} \Big(\mathbb{I}\otimes\overline{V} + R\otimes\overline{W} \Big)  \ec^{(\LL-x)T} \Big] +  \frac{U_{0}}{2} \Tr \Big[ \Big( BR\otimes\overline{BW} + RB\otimes\overline{WB} \Big) \ec^{\LL T} \Big] \nonumber \\ 
& +\int_{0}^{\LL}  \d x  \; \Tr \Big[ B\otimes\overline{Y} \ec^{xT}  \Big([Q,R]\otimes[\overline{Q},\overline{R}] + c R^{2}\otimes\overline{R}^{2}  - \mu R\otimes\overline{R}\Big)  \ec^{(\LL-x)T} \Big] +  \frac{U_{0}}{2} \Tr \Big[ \Big( BR\otimes\overline{YR} + RB\otimes\overline{RY} \Big) \ec^{\LL T} \Big] \nonumber \\
& +  \int_{0}^{\LL}  \d x  \; \Tr \Big[\Big\{ \frac{1}{\epsilon} \Big(BR - \ec^{\ic2\pi\Omega} RB\Big) \otimes \Big(\overline{BR} - \ec^{-\ic2\pi\Omega} \overline{RB}\Big) \nonumber \\
&\quad\quad\quad\quad\quad\; +\frac{1}{2}\Big(B[Q,R] + \ec^{\ic2\pi\Omega} [Q,R]B \Big) \otimes \Big(\overline{BR} - \ec^{-\ic2\pi\Omega} \overline{RB}\Big) \nonumber \\ 
&\quad\quad\quad\quad\quad\; +\frac{1}{2} \Big(BR - \ec^{\ic2\pi\Omega} RB\Big) \otimes \Big( \overline{B[Q,R]} +  \ec^{-\ic2\pi\Omega}\overline{[Q,R]B} \Big) \Big\} \ec^{xT} \Big(\mathbb{I}\otimes\overline{V} + R\otimes\overline{W} \Big)  \ec^{(\LL-x)T} \Big] \nonumber \\
&+  \frac{1}{\epsilon} \Tr \Big[ \Big(BR - \ec^{\ic2\pi\Omega} RB\Big) \otimes \Big\{ \Big(\overline{BW} - \ec^{-\ic2\pi\Omega} \overline{WB}\Big) +\Big(\overline{YR} - \ec^{-\ic2\pi\Omega} \overline{RY}\Big) \Big\} \ec^{\LL T} \Big]  \nonumber \\
&+  \frac{1}{2} \Tr \Big[ \Big(B[Q,R] + \ec^{\ic2\pi\Omega} [Q,R]B \Big) \otimes \Big\{ \Big(\overline{BW} - \ec^{-\ic2\pi\Omega} \overline{WB}\Big) + \Big(\overline{YR} - \ec^{-\ic2\pi\Omega} \overline{RY}\Big)  \Big\} \ec^{\LL T} \Big]  \nonumber \\
&+  \frac{1}{2} \Tr \Big[ \Big(BR - \ec^{\ic2\pi\Omega} RB\Big) \otimes \Big\{ \Big( \overline{B[Q,W]} +  \ec^{-\ic2\pi\Omega}\overline{[Q,W]B} \Big) + \Big( \overline{B[V,R]} +  \ec^{-\ic2\pi\Omega}\overline{[V,R]B} \Big) \nonumber \\
&\quad\quad\quad\quad\quad\quad\quad\quad\quad\quad\quad\quad\quad\quad\quad\quad\quad\quad\quad\quad\quad\quad\quad\quad\;\;\;\;\;\;+ \Big( \overline{Y[Q,R]} +  \ec^{-\ic2\pi\Omega}\overline{[Q,R]Y} \Big) \Big\} \ec^{\LL T} \Big]   \;.
\end{align}
The integrals can again be computed in the same way as for the Gram matrix by inserting a spectral decomposition of $T$ giving rise to
\begin{align}
&\bra{\Phi[\overline{V},\overline{W},\overline{Y}]}\operator{H}_{\epsilon}\ket{\Psi(Q,R,B)}  \approx   \nonumber \\
& \Big\{ \sum_{i\neq j\neq k = 1}^{m_{1}} \frac{(\lambda_{i} - \lambda_{k})(\ec^{\lambda_{i}\LL} - \ec^{\lambda_{j}\LL}) - (\lambda_{i} - \lambda_{j})(\ec^{\lambda_{i}\LL} - \ec^{\lambda_{k}\LL})}{(\lambda_{i} - \lambda_{j})(\lambda_{i} - \lambda_{k})(\lambda_{j} - \lambda_{k})}  \nonumber  \\
& \qquad\qquad\qquad\times \Tr \Big[ B\otimes\overline{B}\rket{r_{i}}\rbra{l_{i}} \Big([Q,R]\otimes[\overline{Q},\overline{R}] + c R^{2}\otimes\overline{R}^{2}  - \mu R\otimes\overline{R}\Big)  \rket{r_{j}}\rbra{l_{j}}   \Big(\mathbb{I}\otimes\overline{V} + R\otimes\overline{W} \Big)  \rket{r_{k}}\rbra{l_{k}} \Big]  \nonumber \\
&+\left(\sum_{i\neq j = 1}^{m_{2}}  \frac{\LL\ec^{\lambda_{j}\LL}}{(\lambda_{j} - \lambda_{i})} + \sum_{i\neq j = 1}^{m_{1}}  \frac{\ec^{\lambda_{i}\LL} - \ec^{\lambda_{j}\LL}}{(\lambda_{j} - \lambda_{i})^{2}} \right) \nonumber \\
&\qquad\qquad\qquad \times\Tr \Big[ B\otimes\overline{B}\rket{r_{i}}\rbra{l_{i}} \Big([Q,R]\otimes[\overline{Q},\overline{R}] + c R^{2}\otimes\overline{R}^{2}  - \mu R\otimes\overline{R}\Big)  \rket{r_{j}}\rbra{l_{j}}   \Big(\mathbb{I}\otimes\overline{V} + R\otimes\overline{W} \Big)  \rket{r_{j}}\rbra{l_{j}} \nonumber \\
&\qquad\qquad\qquad\quad\;\;+  B\otimes\overline{B}\rket{r_{j}}\rbra{l_{j}} \Big([Q,R]\otimes[\overline{Q},\overline{R}] + c R^{2}\otimes\overline{R}^{2}  - \mu R\otimes\overline{R}\Big)  \rket{r_{i}}\rbra{l_{i}}   \Big(\mathbb{I}\otimes\overline{V} + R\otimes\overline{W} \Big)  \rket{r_{j}}\rbra{l_{j}}  \nonumber \\
&\qquad\qquad\qquad\quad\;\;+  B\otimes\overline{B}\rket{r_{j}}\rbra{l_{j}} \Big([Q,R]\otimes[\overline{Q},\overline{R}] + c R^{2}\otimes\overline{R}^{2}  - \mu R\otimes\overline{R}\Big) \rket{r_{j}}\rbra{l_{j}}   \Big(\mathbb{I}\otimes\overline{V} + R\otimes\overline{W} \Big)  \rket{r_{i}}\rbra{l_{i}} \Big] \nonumber \\
&+\frac{\LL^{2}}{2}\sum_{i = 1}^{m3} \ec^{\lambda_{i}\LL} \Tr \Big[ B\otimes\overline{B}\rket{r_{i}}\rbra{l_{i}}\Big([Q,R]\otimes[\overline{Q},\overline{R}] + c R^{2}\otimes\overline{R}^{2}  - \mu R\otimes\overline{R}\Big)  \rket{r_{i}}\rbra{l_{i}}   \Big(\mathbb{I}\otimes\overline{V} + R\otimes\overline{W} \Big)  \rket{r_{i}}\rbra{l_{i}} \Big]  \nonumber \\
 \nonumber \\
&+\sum_{i\neq j = 1}^{m_{1}}  \Tr \Big[ B\otimes\overline{B}\rket{r_{i}}\rbra{l_{i}} \Big([Q,R]\otimes[\overline{Q},\overline{R}] + c R^{2}\otimes\overline{R}^{2}  - \mu R\otimes\overline{R}\Big) \rket{r_{j}}\rbra{l_{j}} \nonumber \\
&\qquad\qquad\qquad\qquad\qquad\qquad\qquad\qquad\qquad \times \Big(\mathbb{I}\otimes\overline{V} + R\otimes\overline{W} \Big) \Big( \frac{\ec^{\lambda_{i}\LL}}{(\lambda_{i} - \lambda_{j})}  \Big(\tilde{T}_{i}^{m_{1}}\Big)^{\rm{p}} - \frac{\ec^{\lambda_{j}\LL}}{(\lambda_{i} - \lambda_{j})}  \Big(\tilde{T}_{j}^{m_{1}}\Big)^{\rm{p}} \Big)  \nonumber \\
&+\sum_{i\neq k = 1}^{m_{1}}  \Tr \Big[ B\otimes\overline{B}\rket{r_{i}}\rbra{l_{i}} \Big([Q,R]\otimes[\overline{Q},\overline{R}] + c R^{2}\otimes\overline{R}^{2}  - \mu R\otimes\overline{R}\Big)    \nonumber \\
&\qquad\qquad\qquad\qquad\qquad\qquad\qquad\qquad\qquad \times \Big( \frac{\ec^{\lambda_{i}\LL}}{(\lambda_{i} - \lambda_{k})}  \Big(\tilde{T}_{i}^{m_{1}}\Big)^{\rm{p}} - \frac{\ec^{\lambda_{k}\LL}}{(\lambda_{i} - \lambda_{k})}  \Big(\tilde{T}_{k}^{m_{1}}\Big)^{\rm{p}} \Big)   \Big(\mathbb{I}\otimes\overline{V} + R\otimes\overline{W} \Big) \rket{r_{k}}\rbra{l_{k}} \Big]  \nonumber \\
&+\sum_{j\neq k = 1}^{m_{1}}  \Tr \Big[ B\otimes\overline{B} \Big( \frac{\ec^{\lambda_{j}\LL}}{(\lambda_{j} - \lambda_{k})}  \Big(\tilde{T}_{j}^{m_{1}}\Big)^{\rm{p}} - \frac{\ec^{\lambda_{k}\LL}}{(\lambda_{j} - \lambda_{k})}  \Big(\tilde{T}_{k}^{m_{1}}\Big)^{\rm{p}} \Big) \Big([Q,R]\otimes[\overline{Q},\overline{R}] + c R^{2}\otimes\overline{R}^{2}  - \mu R\otimes\overline{R}\Big) \nonumber \\
 &\qquad\qquad\qquad\qquad\qquad\qquad\qquad\qquad\qquad \times\rket{r_{j}}\rbra{l_{j}}    \Big(\mathbb{I}\otimes\overline{V} + R\otimes\overline{W} \Big) \rket{r_{k}}\rbra{l_{k}} \Big] \nonumber \\
& \nonumber  \\
&+\sum_{i= 1}^{m_{1}} \ec^{\lambda_{i}\LL} \Tr \Big[ B\otimes\overline{B}\rket{r_{i}}\rbra{l_{i}} \Big([Q,R]\otimes[\overline{Q},\overline{R}] + c R^{2}\otimes\overline{R}^{2}  - \mu R\otimes\overline{R}\Big)   \Big(\tilde{T}_{i}^{m_{1}}\Big)^{\rm{p}}   \Big(\mathbb{I}\otimes\overline{V} + R\otimes\overline{W} \Big)  \Big(\tilde{T}_{i}^{m_{1}}\Big)^{\rm{p}} \Big]  \nonumber \\
 &+\sum_{j = 1}^{m_{1}} \ec^{\lambda_{j}\LL}  \Tr \Big[ B\otimes\overline{B} \Big(\tilde{T}_{j}^{m_{1}}\Big)^{\rm{p}} \Big([Q,R]\otimes[\overline{Q},\overline{R}] + c R^{2}\otimes\overline{R}^{2}  - \mu R\otimes\overline{R}\Big) \rket{r_{j}}\rbra{l_{j}}    \Big(\mathbb{I}\otimes\overline{V} + R\otimes\overline{W} \Big)  \Big(\tilde{T}_{j}^{m_{1}}\Big)^{\rm{p}} \Big]  \nonumber \\
&+\sum_{k = 1}^{m_{1}} \ec^{\lambda_{k}\LL} \Tr \Big[ B\otimes\overline{B} \Big(\tilde{T}_{k}^{m_{1}}\Big)^{\rm{p}} \Big([Q,R]\otimes[\overline{Q},\overline{R}] + c R^{2}\otimes\overline{R}^{2}  - \mu R\otimes\overline{R}\Big) \Big(\tilde{T}_{k}^{m_{1}}\Big)^{\rm{p}}  \Big(\mathbb{I}\otimes\overline{V} + R\otimes\overline{W} \Big)  \rket{r_{k}}\rbra{l_{k}}   \Big]  \nonumber \\
&  \nonumber \\
&-\sum_{i = 1}^{m_{1}} \ec^{\lambda_{i}\LL} \Tr \Big[ B\otimes\overline{B} \Big(\hat{T}_{i}^{m_{1}}\Big)^{\rm{p}} \Big([Q,R]\otimes[\overline{Q},\overline{R}] + c R^{2}\otimes\overline{R}^{2}  - \mu R\otimes\overline{R}\Big) \rket{r_{i}}\rbra{l_{i}}  \Big(\mathbb{I}\otimes\overline{V} + R\otimes\overline{W} \Big)  \rket{r_{i}}\rbra{l_{i}}   \Big]  \nonumber \\
&-\sum_{i = 1}^{m_{1}} \ec^{\lambda_{i}\LL} \Tr \Big[ B\otimes\overline{B} \rket{r_{i}}\rbra{l_{i}} \Big([Q,R]\otimes[\overline{Q},\overline{R}] + c R^{2}\otimes\overline{R}^{2}  - \mu R\otimes\overline{R}\Big) \Big(\hat{T}_{i}^{m_{1}}\Big)^{\rm{p}}  \Big(\mathbb{I}\otimes\overline{V} + R\otimes\overline{W} \Big)  \rket{r_{i}}\rbra{l_{i}}   \Big]  \nonumber \\
&-\sum_{i = 1}^{m_{1}} \ec^{\lambda_{i}\LL} \Tr \Big[ B\otimes\overline{B} \rket{r_{i}}\rbra{l_{i}} \Big([Q,R]\otimes[\overline{Q},\overline{R}] + c R^{2}\otimes\overline{R}^{2}  - \mu R\otimes\overline{R}\Big)  \rket{r_{i}}\rbra{l_{i}} \Big(\mathbb{I}\otimes\overline{V} + R\otimes\overline{W} \Big) \Big(\hat{T}_{i}^{m_{1}}\Big)^{\rm{p}}   \Big]  \nonumber \\
&  \nonumber \\ 
&+L\sum_{i = 1}^{m_{2}} \ec^{\lambda_{i}\LL} \Tr \Big[ B\otimes\overline{B} \Big(\tilde{T}_{i}^{m_{2}}\Big)^{\rm{p}} \Big([Q,R]\otimes[\overline{Q},\overline{R}] + c R^{2}\otimes\overline{R}^{2}  - \mu R\otimes\overline{R}\Big)  \rket{r_{i}}\rbra{l_{i}}  \Big(\mathbb{I}\otimes\overline{V} + R\otimes\overline{W} \Big)  \rket{r_{i}}\rbra{l_{i}}   \Big]  \nonumber \\
&+L\sum_{i = 1}^{m_{2}} \ec^{\lambda_{i}\LL} \Tr \Big[ B\otimes\overline{B} \rket{r_{i}}\rbra{l_{i}} \Big([Q,R]\otimes[\overline{Q},\overline{R}] + c R^{2}\otimes\overline{R}^{2}  - \mu R\otimes\overline{R}\Big) \Big(\tilde{T}_{i}^{m_{2}}\Big)^{\rm{p}}  \Big(\mathbb{I}\otimes\overline{V} + R\otimes\overline{W} \Big)  \rket{r_{i}}\rbra{l_{i}}   \Big]  \nonumber \\
&+L\sum_{i = 1}^{m_{2}} \ec^{\lambda_{i}\LL} \Tr \Big[ B\otimes\overline{B} \rket{r_{i}}\rbra{l_{i}} \Big([Q,R]\otimes[\overline{Q},\overline{R}] + c R^{2}\otimes\overline{R}^{2}  - \mu R\otimes\overline{R}\Big)  \rket{r_{i}}\rbra{l_{i}} \Big(\mathbb{I}\otimes\overline{V} + R\otimes\overline{W} \Big) \Big(\tilde{T}_{i}^{m_{2}}\Big)^{\rm{p}}   \Big] \nonumber \\
& + \Big([Q,R]\otimes[\overline{Q},\overline{R}] + c R^{2}\otimes\overline{R}^{2}  - \mu R\otimes\overline{R}\Big) \leftrightarrow\Big(\mathbb{I}\otimes\overline{V} + R\otimes\overline{W} \Big)  \Big\} \nonumber \\
& \nonumber \\
& + \sum_{i\neq j = 1}^{m_{1}}  \frac{\ec^{\lambda_{i}\LL} - \ec^{\lambda_{j}\LL}}{(\lambda_{i} - \lambda_{j})} \Tr \Big[ B\otimes\overline{B}\rket{r_{i}}\rbra{l_{i}} \Big([Q,R]\otimes([\overline{Q},\overline{W}] + [\overline{V},\overline{R}]) + cR^{2}\otimes(\overline{R}\overline{W} + \overline{W}\overline{R}) -\mu R\otimes\overline{W} \Big)  \rket{r_{j}}\rbra{l_{j}} \Big]  \nonumber \\
&+ \sum_{i = 1}^{m_{1}} \ec^{\lambda_{i}\LL}   \Tr \Big[ B\otimes\overline{B} \rket{r_{i}}\rbra{l_{i}} \Big([Q,R]\otimes([\overline{Q},\overline{W}] + [\overline{V},\overline{R}]) + cR^{2}\otimes(\overline{R}\overline{W} + \overline{W}\overline{R}) -\mu R\otimes\overline{W} \Big) \Big(\tilde{T}_{i}^{m_{1}}\Big)^{\rm{p}} \nonumber \\
&\qquad\qquad\;\;\;+ B\otimes\overline{B} \Big(\tilde{T}_{i}^{m_{1}}\Big)^{\rm{p}}  \Big([Q,R]\otimes([\overline{Q},\overline{W}] + [\overline{V},\overline{R}]) + cR^{2}\otimes(\overline{R}\overline{W} + \overline{W}\overline{R}) -\mu R\otimes\overline{W} \Big) \rket{r_{i}}\rbra{l_{i}} \Big] \nonumber \\
& + \LL \sum_{i=1}^{m_{2}} \ec^{\lambda_{i} \LL}  \Tr \Big[ B\otimes\overline{B}\rket{r_{i}}\rbra{l_{i}}  \Big([Q,R]\otimes([\overline{Q},\overline{W}] + [\overline{V},\overline{R}]) + cR^{2}\otimes(\overline{R}\overline{W} + \overline{W}\overline{R}) -\mu R\otimes\overline{W} \Big) \rket{r_{i}}\rbra{l_{i}} \Big] \nonumber \\
& + \frac{U_{0}}{2} \sum_{i\neq j = 1}^{m_{1}}  \frac{\ec^{\lambda_{i}\LL} - \ec^{\lambda_{j}\LL}}{(\lambda_{i} - \lambda_{j})}  \Tr \Big[ \Big( BR\otimes\overline{BR} + RB\otimes\overline{RB} \Big) \rket{r_{i}}\rbra{l_{i}}  \Big(\mathbb{I}\otimes\overline{V} + R\otimes\overline{W} \Big) \rket{r_{j}}\rbra{l_{j}} \Big]  \nonumber \\
&+  \frac{U_{0}}{2} \sum_{i = 1}^{m_{1}} \ec^{\lambda_{i}\LL}   \Tr \Big[  \Big( BR\otimes\overline{BR} + RB\otimes\overline{RB} \Big)  \rket{r_{i}}\rbra{l_{i}}  \Big(\mathbb{I}\otimes\overline{V} + R\otimes\overline{W} \Big) \Big(\tilde{T}_{i}^{m_{1}}\Big)^{\rm{p}} \nonumber \\
&\qquad\qquad\;\;\;+  \Big( BR\otimes\overline{BR} + RB\otimes\overline{RB} \Big) \Big(\tilde{T}_{i}^{m_{1}}\Big)^{\rm{p}} \Big(\mathbb{I}\otimes\overline{V} + R\otimes\overline{W} \Big) \rket{r_{i}}\rbra{l_{i}} \Big] \nonumber \\
& + \LL \frac{U_{0}}{2} \sum_{i=1}^{m_{2}} \ec^{\lambda_{i} \LL}  \Tr \Big[  \Big( BR\otimes\overline{BR} + RB\otimes\overline{RB} \Big) \rket{r_{i}}\rbra{l_{i}}   \Big(\mathbb{I}\otimes\overline{V} + R\otimes\overline{W} \Big)  \rket{r_{i}}\rbra{l_{i}} \Big] \nonumber \\
& + \frac{U_{0}}{2} \sum_{i=1}^{m_{1}} \ec^{\lambda_{i} \LL}  \Tr \Big[   \Big( BR\otimes\overline{BW} + RB\otimes\overline{WB} \Big) \rket{r_{i}}\rbra{l_{i}}  \Big] \nonumber \\
& + \sum_{i\neq j = 1}^{m_{1}}  \frac{\ec^{\lambda_{i}\LL} - \ec^{\lambda_{j}\LL}}{(\lambda_{i} - \lambda_{j})} \Tr \Big[ B\otimes\overline{Y}\rket{r_{i}}\rbra{l_{i}} \Big([Q,R]\otimes[\overline{Q},\overline{R}] + c R^{2}\otimes\overline{R}^{2}  - \mu R\otimes\overline{R}\Big)  \rket{r_{j}}\rbra{l_{j}} \Big]  \nonumber \\
&+ \sum_{i = 1}^{m_{1}} \ec^{\lambda_{i}\LL}   \Tr \Big[ B\otimes\overline{Y} \rket{r_{i}}\rbra{l_{i}} \Big([Q,R]\otimes[\overline{Q},\overline{R}] + c R^{2}\otimes\overline{R}^{2}  - \mu R\otimes\overline{R}\Big)  \Big(\tilde{T}_{i}^{m_{1}}\Big)^{\rm{p}} \nonumber \\
&\qquad\qquad\;\;\;+ B\otimes\overline{Y} \Big(\tilde{T}_{i}^{m_{1}}\Big)^{\rm{p}} \Big([Q,R]\otimes[\overline{Q},\overline{R}] + c R^{2}\otimes\overline{R}^{2}  - \mu R\otimes\overline{R}\Big)  \rket{r_{i}}\rbra{l_{i}} \Big] \nonumber \\
& + \frac{U_{0}}{2} \sum_{i=1}^{m_{1}} \ec^{\lambda_{i} \LL}  \Tr \Big[   \Big( BR\otimes\overline{YR} + RB\otimes\overline{RY} \Big) \rket{r_{i}}\rbra{l_{i}}  \Big] \nonumber \\
& + \sum_{i\neq j = 1}^{m_{1}}  \frac{\ec^{\lambda_{i}\LL} - \ec^{\lambda_{j}\LL}}{(\lambda_{i} - \lambda_{j})} \Tr \Big[\Big\{ \frac{1}{\epsilon} \Big(BR - \ec^{\ic2\pi\Omega} RB\Big) \otimes \Big(\overline{BR} - \ec^{-\ic2\pi\Omega} \overline{RB}\Big) \nonumber \\
&\;\;\;\;\quad\quad\quad\quad\quad+\frac{1}{2}\Big(B[Q,R] + \ec^{\ic2\pi\Omega} [Q,R]B \Big) \otimes \Big(\overline{BR} - \ec^{-\ic2\pi\Omega} \overline{RB}\Big) \nonumber \\ 
&\;\;\;\;\quad\quad\quad\quad\quad+\frac{1}{2} \Big(BR - \ec^{\ic2\pi\Omega} RB\Big) \otimes \Big( \overline{B[Q,R]} +  \ec^{-\ic2\pi\Omega}\overline{[Q,R]B} \Big)\Big\} \rket{r_{i}}\rbra{l_{i}}  \Big(\mathbb{I}\otimes\overline{V} + R\otimes\overline{W} \Big) \rket{r_{j}}\rbra{l_{j}} \Big]  \nonumber \\
& + \sum_{i=1}^{m_{1}} \ec^{\lambda_{i}\LL} \Tr \Big[\Big\{ \frac{1}{\epsilon} \Big(BR - \ec^{\ic2\pi\Omega} RB\Big) \otimes \Big(\overline{BR} - \ec^{-\ic2\pi\Omega} \overline{RB}\Big) \nonumber \\
&\;\;\;\;\quad\quad\quad\quad+\frac{1}{2}\Big(B[Q,R] + \ec^{\ic2\pi\Omega} [Q,R]B \Big) \otimes \Big(\overline{BR} - \ec^{-\ic2\pi\Omega} \overline{RB}\Big) \nonumber \\ 
&\;\;\;\;\quad\quad\quad\quad+\frac{1}{2} \Big(BR - \ec^{\ic2\pi\Omega} RB\Big) \otimes \Big( \overline{B[Q,R]} +  \ec^{-\ic2\pi\Omega}\overline{[Q,R]B} \Big) \Big\} \Big\{ \rket{r_{i}}\rbra{l_{i}}  \Big(\mathbb{I}\otimes\overline{V} + R\otimes\overline{W} \Big)  \Big(\tilde{T}_{i}^{m_{1}}\Big)^{\rm{p}} \nonumber \\
&\;\quad\quad\quad\quad\quad\quad\quad\quad\quad\quad\quad\quad\quad\quad\quad\quad\quad\quad\quad\quad\quad\quad\quad\quad\quad\quad\quad+ \Big(\tilde{T}_{i}^{m_{1}}\Big)^{\rm{p}}  \Big(\mathbb{I}\otimes\overline{V} + R\otimes\overline{W} \Big) \rket{r_{i}}\rbra{l_{i}} \Big\} \Big] \nonumber \\
&+ \LL \sum_{i=1}^{m_{2}} \ec^{\lambda_{i}\LL} \Tr \Big[\Big\{ \frac{1}{\epsilon} \Big(BR - \ec^{\ic2\pi\Omega} RB\Big) \otimes \Big(\overline{BR} - \ec^{-\ic2\pi\Omega} \overline{RB}\Big) \nonumber \\
&\;\;\;\;\quad\quad\quad\quad+\frac{1}{2}\Big(B[Q,R] + \ec^{\ic2\pi\Omega} [Q,R]B \Big) \otimes \Big(\overline{BR} - \ec^{-\ic2\pi\Omega} \overline{RB}\Big) \nonumber \\ 
&\;\;\;\;\quad\quad\quad\quad+\frac{1}{2} \Big(BR - \ec^{\ic2\pi\Omega} RB\Big) \otimes \Big( \overline{B[Q,R]} +  \ec^{-\ic2\pi\Omega}\overline{[Q,R]B} \Big)  \Big\} \rket{r_{i}}\rbra{l_{i}}  \Big(\mathbb{I}\otimes\overline{V} + R\otimes\overline{W} \Big)  \rket{r_{i}}\rbra{l_{i}} \Big] \nonumber \\
&+ \sum_{i=1}^{m_{1}} \ec^{\lambda_{i}\LL}  \frac{1}{\epsilon} \Tr \Big[ \Big(BR - \ec^{\ic2\pi\Omega} RB\Big) \otimes \Big\{ \Big(\overline{BW} - \ec^{-\ic2\pi\Omega} \overline{WB}\Big) +\Big(\overline{YR} - \ec^{-\ic2\pi\Omega} \overline{RY}\Big) \Big\} \rket{r_{i}}\rbra{l_{i}} \Big]  \nonumber \\
&+ \sum_{i=1}^{m_{1}} \ec^{\lambda_{i}\LL}  \frac{1}{2} \Tr \Big[ \Big(B[Q,R] + \ec^{\ic2\pi\Omega} [Q,R]B \Big) \otimes \Big\{ \Big(\overline{BW} - \ec^{-\ic2\pi\Omega} \overline{WB}\Big) + \Big(\overline{YR} - \ec^{-\ic2\pi\Omega} \overline{RY}\Big)  \Big\} \rket{r_{i}}\rbra{l_{i}} \Big]  \nonumber \\
&+ \sum_{i=1}^{m_{1}} \ec^{\lambda_{i}\LL}  \frac{1}{2} \Tr \Big[ \Big(BR - \ec^{\ic2\pi\Omega} RB\Big) \otimes \Big\{ \Big( \overline{B[Q,W]} +  \ec^{-\ic2\pi\Omega}\overline{[Q,W]B} \Big) + \Big( \overline{B[V,R]} +  \ec^{-\ic2\pi\Omega}\overline{[V,R]B} \Big) \nonumber \\
&\quad\quad\quad\quad\quad\quad\quad\quad\quad\quad\quad\quad\quad\quad\quad\quad\quad\quad\quad\quad\quad\quad\quad\quad\;\;\;\;\;\;+ \Big( \overline{Y[Q,R]} +  \ec^{-\ic2\pi\Omega}\overline{[Q,R]Y} \Big) \Big\} \rket{r_{i}}\rbra{l_{i}} \Big]  \;.
\end{align}
The ordering of the terms in the above expression goes along the same lines as for the calculation of the Gram matrix. Moreover, the abbreviation $(...)\leftrightarrow(...)$ has to be understood as the sum of all the terms inside the corresponding curly brackets but with the two brackets exchanged. Note that in principle one can choose a different (smaller) \textit{tol} in the computation of the gradient as for the computation of the Gram matrix which leads to different (larger) integers $m_{1},m_{2}$ and $m_{3}$ respectively. 

To conclude, we will calculate the vector $\left[ y_{1};y_{2};y_{3}\right]$ which was introduced in section \ref{sec:secTDVP} of the main paper. The calculation goes along the same lines as for the Gram matrix and the gradient and we find that Eq.~\ref{eq:gsortho} can be approximated by
 \begin{align}
&\Braket{\Phi[\overline{V},\overline{W},\overline{Y}] | \Psi(Q,R,B)} \approx \nonumber \\ 
&\sum_{i=1}^{m_{1}} \ec^{\lambda_{i}\LL} \Tr  \Big[ B\otimes\overline{B}  \big( \tilde{T}_{i}^{m_{1}} \big)^{{\rm{p}}}  \left(\mathbb{I}\otimes\overline{V} + R\otimes\overline{W} \right)\rket{r_{i}} \rbra{l_{i}} +  \rket{r_{i}} \rbra{l_{i}} \left(\mathbb{I}\otimes\overline{V} + R\otimes\overline{W} \right) \big( \tilde{T}_{i}^{m_{1}} \big)^{{\rm{p}}}   B\otimes\overline{B}\Big] \\
&+ \sum_{\ic\neq k=1}^{m_{1}} \frac{\ec^{\lambda_{i}\LL} - \ec^{\lambda_{k}\LL}}{\lambda_{i}-\lambda_{k}} \Tr \Big[ B\otimes\overline{B} \rket{r_{i}}\rbra{l_{i}} \left(\mathbb{I}\otimes\overline{V} + R\otimes\overline{W} \right)\rket{r_{k}}\rbra{l_{k}}    \Big] \\
&+ \LL\sum_{i=1}^{m_{2}}  \ec^{\lambda_{i}\LL} \Tr  \Big[ B\otimes\overline{B} \rket{r_{i}}\rbra{l_{i}} \left(\mathbb{I}\otimes\overline{V} + R\otimes\overline{W} \right)\rket{r_{i}}\rbra{l_{i}} \Big] \;.
\end{align}   
\\
\\

\section{Scaling behaviour of the ground state energy and finite $D$ scaling of the density profiles}\label{sec:AppB}  

Next we will discus the dependence of the ground state energy of Hamiltonian \eqref{eq:hamileps} on the cutoff $\epsilon$ around position $x=0$. The Hamiltonain is thus split into two parts, $\hat{H}_{\epsilon} = \hat{H}_{\rm{bulk}} + \hat{H}_{\rm{boundary}}$, where the \textit{bulk} part equals to the first line in  \eqref{eq:hamileps} and the \textit{boundary} term consists of the remaining three lines in  \eqref{eq:hamileps}. We are interested in the scaling of the respective ground state energies $E_{\rm{bulk}}(\epsilon)$ and $E_{\rm{boundary}}(\epsilon)$ at a fixed bond dimension $D$.
Fig.(5) shows that both converge linearly with $\epsilon$, with quadratic corrections in $E_{\rm{bulk}}(\epsilon)$ (left panel) not visible in the plotted region for $E_{\rm{boundary}}(\epsilon)$ (right panel).
The asymptotic value $\lim_{\epsilon\rightarrow 0}  E_{\rm{boundary}}(\epsilon)$ decreases with increasing $D$ and is found to be $\mathcal{O}(10^{-6})$ already at $D=16$. 
In the main text we have chosen $\epsilon=2.5\times 10^{-3}$ which corresponds to an error in the energy density of the bulk of roughly $10^{-8}$ (in comparison to $\lim_{\epsilon\rightarrow 0}  E_{\rm{bulk}}(\epsilon)/L$).
A sensitive measure for how well the twisted boundary conditions are imposed is given by $||BR - \ec^{ \ic2\pi\Omega}RB||_{\rm{cMPS}}$, where the norm $||.||_{\rm{cMPS}}$  is weighted by the proper cMPS metric given by $\ec^{\LL T}$. This is precisely the cMPS expectation value of $(\psi(0)-\ec^{ \ic2\pi\Omega}\psi(\LL))^{\dagger}(\psi(0)-\ec^{ \ic2\pi\Omega}\psi(\LL))$ and thus has to decrease linearly for small enough $\epsilon$ as can be seen from the right panel of Fig.(5). In fact this value is $\mathcal{O}(10^{-8})$ for $\epsilon=2.5\times 10^{-3}$ indicating that the twisted boundary conditions are very well approximated even for relatively large $\epsilon>0$.
To conclude, it is perfectly well justified to consider only $E_{\rm{bulk}}(\epsilon)$ as the true approximation to the exact ground state energy as long as $\epsilon$ is chosen small enough.   

 \begin{figure*}\label{fig:escaling}
\includegraphics[width=80mm,height=55mm]{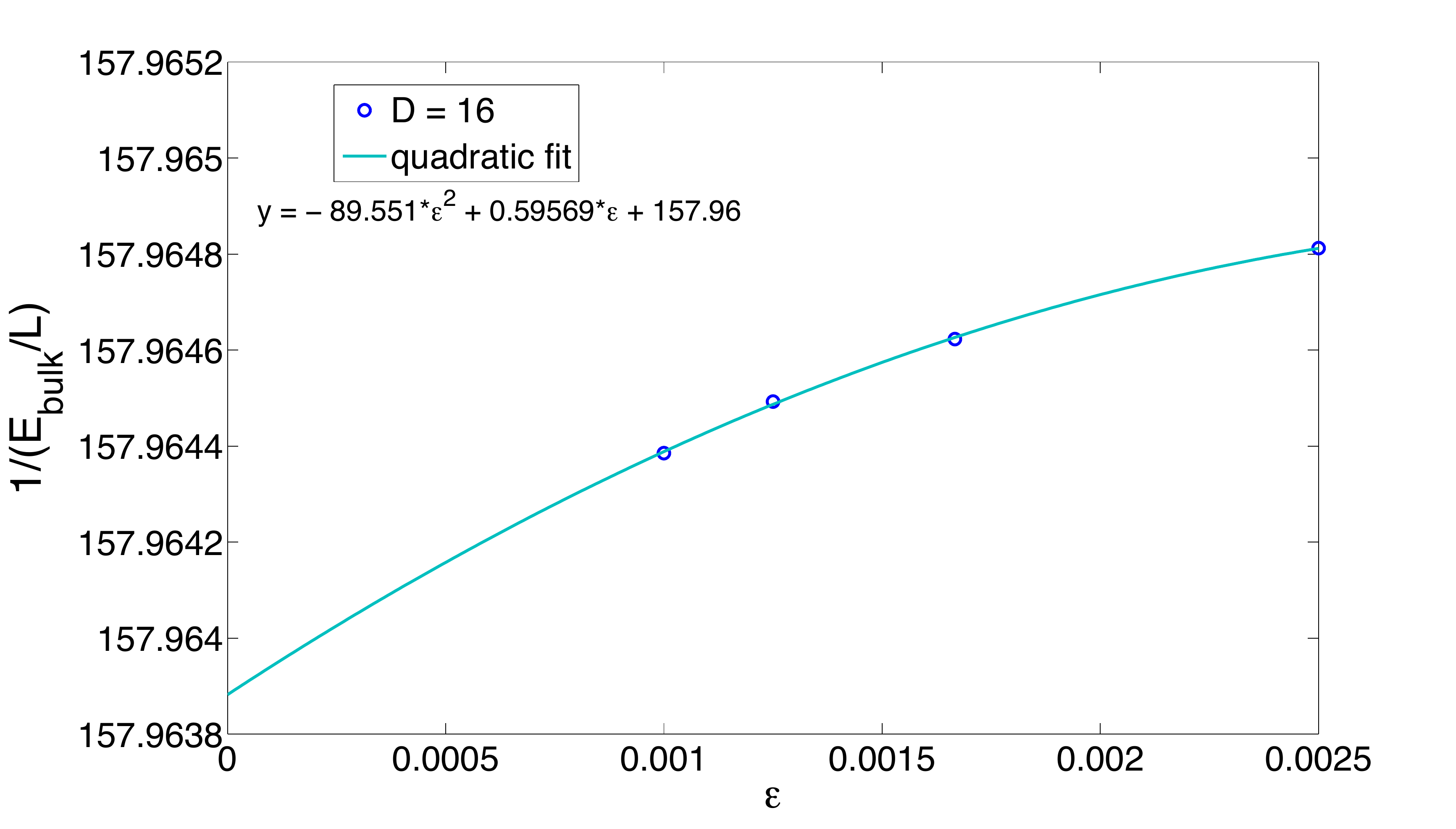}
\includegraphics[width=80mm,height=55mm]{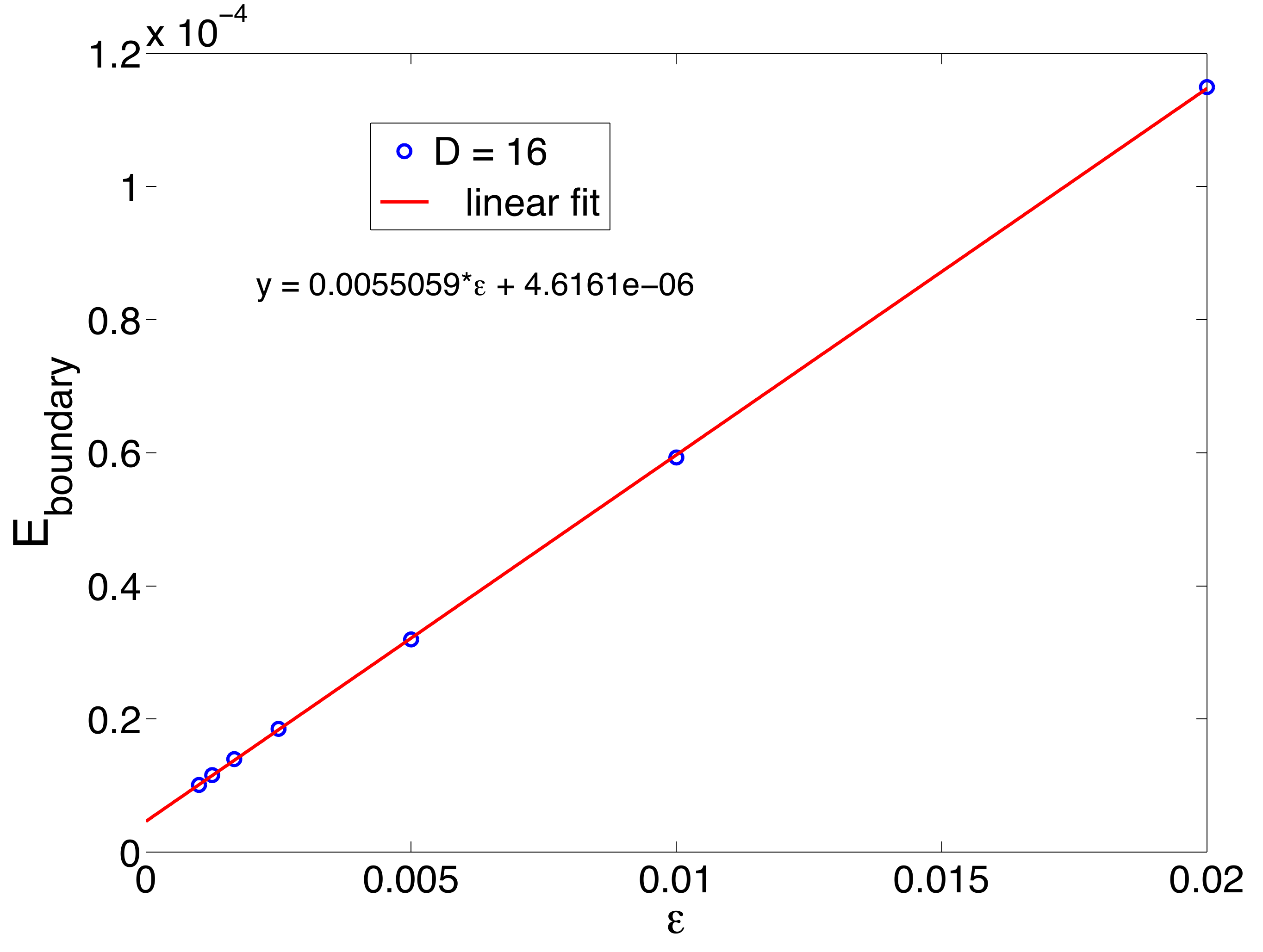}
\caption{\small{Left: Inverse ground state energy density of the bulk as a function of the cutoff $\epsilon$. Right: Ground state expectation value of the boundary terms in Hamiltonain \eqref{eq:hamileps} for several values of $\epsilon$. The other parameters in both panels are $D=16$, $L=256$, $N=32$, $\gamma=80$, $\Lambda=4$ and $\Omega=0$.}}
 \end{figure*}

Finally, we are interested in the convergence behaviour of the density profiles with increasing bond dimension. In Fig.(6) we show density profiles for three different values of $D$ while having all other parameters fixed. As mentioned in section \ref{sec:secResults} we observe that the profiles converge from inside to the outside as the effective cMPS correlation length increases. The central depletion has already almost converged at the lowest value of $D$ while for this value the behaviour in the bulk is still pretty off as can be seen from the strongly damped Friedel oscillations in the inset of Fig.(6) in comparison to predictions of Luttinger liquid theory for the Lieb-Liniger model with obc \cite{0953-4075-37-7-051}.

  \begin{figure*}\label{fig:dscaling}
  \includegraphics[width=100mm,height=60mm]{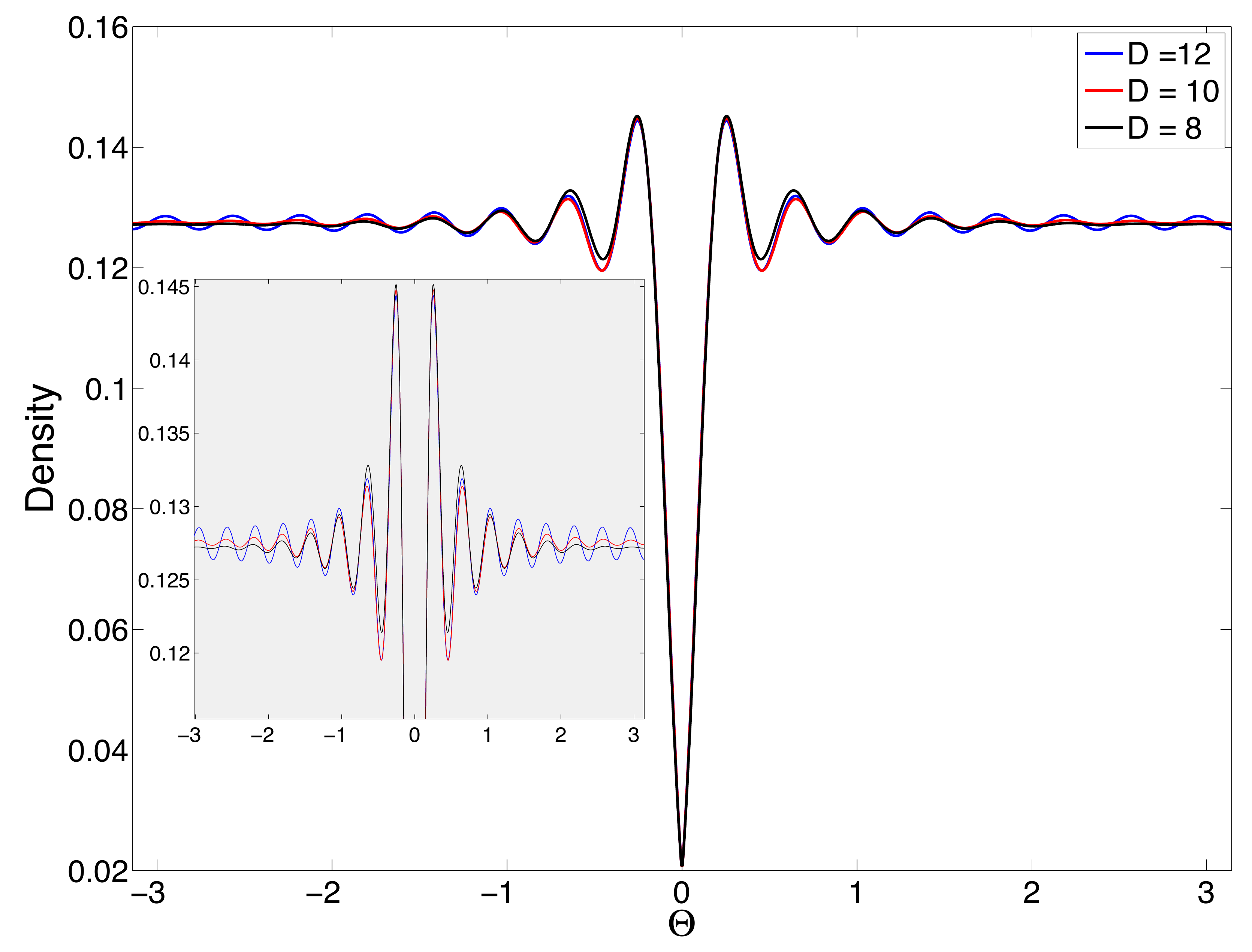}
\caption{\small{Density profile for three different values of $D$ at $L=128$, $N=16$, $\gamma=80$, $\Lambda=4$ and $\Omega=0$. The inset shows a zoom around the bulk value of the density.}}
 \end{figure*}

\end{widetext}
\end{document}